\documentclass[preprint,review,12pt]{elsarticle}

\usepackage{graphics}

\usepackage{amssymb}
\usepackage{subeqn}

\biboptions{sort&compress}

\journal{}

\begin{document}

\begin{frontmatter}



\title{A general multiple-relaxation-time lattice Boltzmann model for nonlinear anisotropic convection-diffusion equations}


\author[a,b]{Zhenhua Chai}
\author{Baochang Shi\fnref{a,b}\corref{cor1}}\cortext[cor1]{Corresponding author.
\hspace*{20pt}}\ead{shibc@hust.edu.cn}
\author[b]{Zhaoli Guo}

\address[a]{School of Mathematics and Statistics, Huazhong University of Science and Technology, Wuhan, 430074, China}
\address[b]{State Key Laboratory of Coal Combustion, Huazhong University of Science and Technology, Wuhan 430074, China}

\begin{abstract}

In this paper, based on the previous work [B. Shi, Z. Guo, Lattice Boltzmann model for nonlinear convection-diffusion equations, Phys. Rev. E 79 (2009) 016701], we develop a general multiple-relaxation-time (MRT) lattice
Boltzmann model for nonlinear anisotropic
convection-diffusion equation (NACDE), and show that the NACDE can be recovered correctly from the present model through the Chapman-Enskog analysis. We then test the MRT model through some
classic CDEs, and find that the numerical
results are in good agreement with analytical solutions or some
available results. Besides, the numerical results also show that similar to the single-relaxation-time (SRT) lattice Boltzmann model or so-called BGK model, the present MRT model also has a second-order convergence rate in space. Finally, we also perform a comparative study on the accuracy and stability of the MRT model and BGK model by using two examples. In terms of the accuracy, both the theoretical analysis and numerical results show that a \emph{numerical} slip on the boundary would be caused in the BGK model, and cannot be eliminated unless the relaxation parameter is fixed to be a special value, while the \emph{numerical} slip in the MRT model can be overcome once the relaxation parameters satisfy some constrains. The results in terms of stability also demonstrate that the MRT model could be more stable than the BGK model through tuning the free relaxation parameters.

\end{abstract}


\end{frontmatter}

%
%

\section{Introduction}

In the past decades, the lattice Boltzmann (LB) method, as a
numerical approach originated from the lattice gas automata or
developed from the simplified kinetic models, has gained great
success in the simulation of the complex flows \cite{Chen, Succi, Aidun, Guo2013},
and has also been extended to solve some partial
differential equations, including
the diffusion equation \cite{Wolf-Gladrow, Huber}, wave equation
\cite{Yan}, Burgers equation \cite{Yu}, Poisson equation \cite{Chai2008a}, isotropic and anisotropic convection-diffusion equations (CDEs) \cite{Dawson, Guo1999, Sman, He, Zhang, Deng, Suga, Zheng, Shi2008,
Chopard, Huang2011, Chai2013, Perko, Ginzburg2005a, Ginzburg2005b,
Ginzburg2012, Servan-Camas, Rasin, Yoshida2010, Yoshida2014, Li,
Huang2014, Chai2014, Liang, Huang2015, Shi2009, Shi2011}. However, most of these
available works related to CDEs mainly focused on the linear CDE for its
important role in the study of the heat and mass transfer
\cite{Dawson, Guo1999, Sman, He, Zhang, Deng, Suga, Zheng, Shi2008,
Chopard, Huang2011, Chai2013, Perko, Ginzburg2005a, Ginzburg2005b,
Ginzburg2012, Servan-Camas, Rasin, Yoshida2010, Yoshida2014, Li,
Huang2014, Chai2014, Liang, Huang2015}. Recently, through constructing a proper equilibrium
distribution function, Shi and Guo \cite{Shi2009} proposed a
single-relaxation-time (SRT) LB model or the
Bhatnagar-Gross-Krook (BGK) model for nonlinear
convection-diffusion equations (NCDEs) where a nonlinear convection and/or diffusion term is included. Although the model has a second-order convergence rate in space \cite{Shi2009, Shi2011}, and can also be considered as an extension to some available works, it still has some limitations. The first is that when the BGK model is used to solve the
NCDEs where the diffusion coefficient is very small, the relaxation parameter would
be close to its limit value, thus the model may suffer from the stability or accuracy problem \cite{Yoshida2010, Chai2014}. Secondly,
the BGK model is usually limited to solve the nonlinear isotropic CDEs since it
does not have sufficient relaxation parameters to describe the
anisotropic diffusion process, while the multiple-relaxation-time (MRT) LB model with more relaxation parameters seems
more suitable and more reasonable in solving NCDE with anisotropic diffusion process. And
finally, it is also well known that in the BGK model, only a single
relaxation process is used to characterize the collision effects,
which means all modes relax to their equilibria with the same rate,
while from a physical point of view, these rates corresponding to
different modes should be different from each other during the
collision process \cite{Guo2013, Lallemand}. To overcome these
deficiencies inherent in the BGK model, in this work we will present
a general MRT LB model for the nonlinear anisotropic CDE (NACDE), and show
that, through the Chapman-Enskog analysis, the NACDE with a source
term can be recovered correctly from the proposed model. Besides, it
is also found that, similar to our previous BGK model \cite{Shi2009,
Shi2011}, the present MRT model also has a second-order convergence
rate in space, while it could be more stable and more accurate than the BGK
model through tuning the relaxation parameters properly.

The rest of the paper is organized as follows. In section 2, a general MRT model for NACDE with a source term is first presented, and then some special cases and distinct characteristics of the present model are also discussed. In section 3, the accuracy and convergence rate of the MRT model are tested through some classic CDEs, followed by a comparison between the present MRT model and the BGK model, and finally, some conclusions are given in section 4.

\section{Multiple-relaxation-time lattice Boltzmann model for nonlinear anisotropic convection-diffusion equations}\label{section2}

\subsection{Nonlinear anisotropic convection-diffusion equation}

The $n$-dimensional ($n$-D) nonlinear anisotropic convection-diffusion equation with a source term can be written as
\begin{equation}\label{eq1}
\partial_{t} \phi+\nabla\cdot\mathbf{B}(\phi)=\nabla\cdot[\mathbf{K}\cdot(\nabla\cdot\mathbf{D}(\phi))]+R(\mathbf{x}, t),
\end{equation}
where $\phi$ is a scalar variable and is a function of time and
space, $\nabla$ is the gradient operator, $\mathbf{K}=\mathbf{K}(\phi , \mathbf{x}, t)$ is the diffusion tensor. $\mathbf{B}(\phi)$ and $\mathbf{D}(\phi)$ are two differential tensor functions with respect to $\phi$, $R(\mathbf{x}, t)$ is
the source term. It should be noted that Eq. (1) can be viewed as a
general form of some important partial differential equations
\cite{Yoshida2010, Shi2009, Shi2011}, such as the (anisotropic) diffusion
equation, Burgers equation, Burgers-Fisher equation, Buckley-Leverett equation, nonlinear heat
conduction equation, (anisotropic) convection-diffusion equation,
and so on.

\subsection{Multiple-relaxation-time lattice Boltzmann model}

Generally, the models in the LB method can be classified into three kinds based on collision operator, i.e., the lattice BGK model \cite{Qian, Ansumali}, the two relaxation-time model \cite{Ginzburg2012, Ginzburg2008}, and the MRT model \cite{Lallemand, dHumieres}. In this work, we will focus on the MRT model for its superiority on the stability and accuracy both in the study of fluid flows \cite{Lallemand, Luo} and solving CDEs \cite{Yoshida2010, Chai2014}.

The MRT model with D$n$Q$q$ lattice ($q$ is the number of discrete
directions) \cite{Qian} for the NACDE [Eq.~(\ref{eq1})] is considered here, and the evolution equation of the model can be written as
\begin{eqnarray}\label{eq2}
\phi_{k}(\mathbf{x} + \mathbf{c}_{k}\delta t,\;t + \delta t) & = &  \phi_{k}(\mathbf{x},\;t )-(\mathbf{M}^{-1}\mathbf{S M})_{kj}[\phi_{j}(\mathbf{x},\;t) - \phi_{j}^{(eq)}(\mathbf{x},\;t)] \nonumber \\
& + & (\delta t+\frac{\delta t^{2}}{2}\bar{D}_{k})R_{k}(\mathbf{x}, t),
\end{eqnarray}
where $\delta t$ is time step, $\bar{D}_{k}=\theta\partial_{t}+\gamma\mathbf{c}_{k}\cdot\nabla$ with $\theta$ and $\gamma$ being two parameters to be determined in the following part. $\phi_{k}(\mathbf{x},\;t)$ is the distribution function associated with the discrete velocity $\mathbf{c}_{k}$ at position $\mathbf{x}$ and time $t$, $\phi_{k}^{(eq)}(\mathbf{x},\;t)$ is the equilibrium distribution function, and is defined as \cite{Shi2009}
\begin{equation}\label{eq3}
\phi_{k}^{(eq)}(\mathbf{x},\;t)=\omega_{k}[\phi+\frac{\mathbf{c}_{k}\cdot \mathbf{B}}{c_{s}^{2}}+\frac{(\mathbf{C}+d c_{s}^{2}\mathbf{D}-c_{s}^{2}\phi\mathbf{I}):(\mathbf{c}_{k}\mathbf{c}_{k}-c_{s}^{2}\mathbf{I})}{2c_{s}^{4}}],
\end{equation}
where $\mathbf{I}$ is the unit matrix, $d$ is a positive parameter related to the diffusion tensor $\mathbf{K}$ [see Eq.~(\ref{eq26})], $\mathbf{c}_{k}$ and $\omega_{k}$ are discrete velocity and weight coefficient, and in different lattice models, they can be defined as

\noindent
\begin{subequations}\label{eq4}
\textrm{D$1$Q$3$: $\;$}
\begin{equation}
\mathbf{c}=(0, 1, -1)c,
\end{equation}
\begin{equation}
\omega_{0}=2/3, \; \omega_{1}=\omega_{2}=1/6,
\end{equation}
\end{subequations}
\begin{subequations}\label{eq5}
\textrm{D$2$Q$9$: $\;$}
\begin{equation}
\mathbf{c}=\left(\begin{array}{ccccccccc}
0 & 1 & 0 & -1 & 0 & 1 & -1 & -1 & 1 \\
0 & 0 & 1 & 0 & -1 & 1 & 1 &-1 & -1
\end{array} \right)c
\end{equation}
\begin{equation}
\omega_{0}=4/9, \; \omega_{k=1-4}=1/9, \; \omega_{k=5-8}=1/36,
\end{equation}
\end{subequations}
\begin{subequations}\label{eq6}
\textrm{D$3$Q$19$: $\;$}
\begin{equation}
\mathbf{c}=\left(\begin{array}{ccccccccccccccccccc}
0 & 1 & -1 & 0 & 0 & 0 & 0 & 1 & -1 & 1 & -1 & 1 & -1 & -1 & 1 & 0 & 0 & 0 & 0\\
0 & 0 & 0 & 1 & -1 & 0 & 0 & 1 & -1 & -1 & 1 & 0 & 0 & 0 & 0 & 1 & -1 & 1 & -1\\
0 & 0 & 0 & 0 & 0 & 1 & -1 & 0 & 0 & 0 & 0 & 1 & -1 & 1 & -1 & 1 & -1 & -1 & 1
\end{array} \right)c
\end{equation}
\begin{equation}
\omega_{0}=1/3, \; \omega_{k=1-6}=1/18, \; \omega_{k=7-18}=1/36,
\end{equation}
\end{subequations}
where $c=\delta x/\delta t$ with $\delta x$ representing lattice spacing, and usually it is not equal to 1, $c_{s}^{2}=c^{2}/3$. To
recover the correct NACDE from present MRT model, the second-order
differential tensor $\mathbf{C}$ in Eq.~(\ref{eq3}) should satisfy
\begin{equation}\label{eq7}
C'_{\alpha\beta}(\phi)=B'_{\alpha}(\phi)B'_{\beta}(\phi)\; \textrm{or}\; C_{\alpha\beta}(\phi)=\int B'_{\alpha}(\phi)B'_{\beta}(\phi) d\phi.
\end{equation}
$\mathbf{M}$ in Eq.~(\ref{eq2}) is a transformation matrix, and can be used to project the distribution function $\phi_{k}$ and equilibrium distribution function $\phi_{k}^{(eq)}$ in the discrete velocity space onto macroscopic variables in the moment space through following relations \cite{Lallemand, Chai2012},
\begin{equation}\label{eq8}
\mathbf{m}:=\mathbf{M\Phi},\ \ \ \mathbf{m}^{(eq)}:=\mathbf{M\Phi}^{(eq)},
\end{equation}
where $\mathbf{\Phi}=(\phi_{0},\cdots, \phi_{q-1})^{\top}$, $\mathbf{\Phi}^{(eq)}=(\phi_{0}^{(eq)},\cdots, \phi_{q-1}^{(eq)})^{\top}$. $\mathbf{S}$ is the relaxation matrix, $R_{k}(\mathbf{x}, t)$ is
the discrete source term, and is defined by
\begin{equation}\label{eq9}
R_{k}(\mathbf{x}, t)=\omega_{k}(1+\frac{\mathbf{c}_{k}\cdot \tilde{\mathbf{B}}}{c_{s}^{2}})R(\mathbf{x}, t),
\end{equation}
where $\tilde{\mathbf{B}}$ is a differential tensor to be determined below.

In addition, to derive correct NACDE, i.e., Eq.~(\ref{eq1}), the following conditions also need to be satisfied,
\begin{subequations}\label{eq10}
\begin{equation}
 \sum\limits_k\phi_{k}  = \sum\limits_{k} \phi_{k}^{(eq)}=\phi,  \ \ \sum\limits_{k}R_{k}  = R,
\end{equation}
\begin{equation}
\sum\limits_{k}\mathbf{c}_{k}\phi_{k}^{(eq)}  = \mathbf{B}(\phi ), \ \ \sum\limits_{k}\mathbf{c}_{k}\mathbf{c}_{k}\phi_{k}^{(eq)}  = \mathbf{C}+d c_{s}^{2}\mathbf{D}(\phi), \ \ \sum\limits_{k}\mathbf{c}_{k}R_{k}  = \tilde{\mathbf{B}}(\phi)R.
\end{equation}
\end{subequations}

\subsection{The Chapman-Enskog analysis}

In this part, we will present the Chapman-Enskog analysis on how to derive the NACDE from the present MRT model. In the Chapman-Enskog analysis, the distribution function, the time and space derivatives, and the source term can be expressed as \cite{Chai2014}
\begin{subequations}\label{eq11}
\begin{equation}
\phi_{k} = \phi_{k}^{(0)} + \epsilon\phi_{k}^{(1)} + \epsilon^{2}\phi_{k}^{(2)} + \cdots,
\end{equation}
\begin{equation}
\partial_{t} = \epsilon\partial _{t_{1}} + \epsilon^{2}\partial_{t_{2}}, \ \ \nabla = \epsilon \nabla_{1}, \ \ R = \epsilon R_{1},
\end{equation}
\end{subequations}
where $\epsilon$ is small parameter.
\noindent Substituting Eqs.~(\ref{eq11}) into Eq.~(\ref{eq2}) and using the Taylor expansion, we can obtain zero, first and second-order equations
in $\epsilon$,
\begin{subequations}\label{eq12}
\begin{equation}
\epsilon^{0}:\ \ \phi_{k}^{(0)} = \phi_{k}^{({eq})},
\end{equation}
\begin{equation}
\epsilon^{1}:\ \  D_{1k}\phi_{k}^{(0)} = -\frac{1}{\delta t}(\mathbf{{M}^{-1}\mathbf{S M}})_{kj}\phi_{j}^{(1)} + R_{k}^{(1)},
\end{equation}
\begin{equation}
\epsilon^{2}:\ \ \partial_{t_2}\phi_{k}^{(0)} + D_{1i}\phi_{k}^{(1)}+\frac{\delta t}{2}D_{1k}^{2}\phi_{k}^{(0)} = -\frac{1}{\delta t}(\mathbf{{M}^{-1}\mathbf{S M}})_{kj}\phi_{j}^{(2)} + \frac{\delta t}{2}\bar{D}_{1k}R_{k}^{(1)},
\end{equation}
\end{subequations}
where $D_{1k}=\partial_{t_1} + \mathbf{c}_{k}\cdot\nabla_{1}$, $\bar{D}_{1k}=\theta\partial_{t_1} + \gamma\mathbf{c}_{k}\cdot\nabla_{1}$. If we rewrite Eqs.~(\ref{eq12}) in a vector form, and multiply $\mathbf{M}$ on both sides of them, we can obtain the corresponding equations in moment space,
\begin{subequations}\label{eq13}
\begin{equation}
\epsilon^{0}:\ \ \mathbf{m}^{(0)} = \mathbf{m}^{({eq})},
\end{equation}
\begin{equation}
\epsilon^{1}:\ \ \mathbf{D}_{1}\mathbf{m}^{(0)} = -\frac{1}{\delta t}\mathbf{S}\mathbf{m}^{(1)} + \mathbf{M R}^{(1)},
\end{equation}
\begin{equation}
\epsilon^{2}:\ \ \partial_{t_2}\mathbf{m}^{(0)} + \mathbf{D}_{1}(\mathbf{I}-\frac{\mathbf{S}}{2})\mathbf{m}^{(1)} = -\frac{1}{\delta t}\mathbf{S}\mathbf{m}^{(2)} + \frac{\delta t}{2}\mathbf{M}\bar{\bar{\mathbf{D}}}_{1}\mathbf{M}\mathbf{R}^{(1)},
\end{equation}
\end{subequations}
where $\mathbf{D}_{1}=\mathbf{I}\ \partial_{t_1}+\mathbf{M}\ \textrm{diag}(c_{0\alpha}, \cdots, c_{(q-1)\alpha})\nabla_{1\alpha}\mathbf{M}^{-1}$, $\bar{\bar{\mathbf{D}}}_{1}=(\theta-1)\mathbf{I}\ \partial_{t_1}+(\gamma-1)\mathbf{M}\ \textrm{diag}(c_{0\alpha}, \cdots, c_{(q-1)\alpha})\nabla_{1\alpha}\mathbf{M}^{-1}$, $\mathbf{R}^{(1)}=(R_{0}^{(1)}, \cdots, R_{q-1}^{(1)})^{\top}$, $\mathbf{m}^{(1)}=\mathbf{M\Phi}^{(1)}$ and $\mathbf{m}^{(2)}=\mathbf{M\Phi}^{(2)}$ with $\mathbf{\Phi}^{(k)}=(\phi_{0}^{(k)}, \cdots, \phi_{q-1}^{(k)})^{\top}$.

If we take the D2Q9 lattice model as an example, the transportation
matrix $\mathbf{M}$ can be written as
\begin{equation}\label{eq14}
\mathbf{M}=\mathbf{C}_{d}\mathbf{M}_{0},
\end{equation}
where ${\mathbf{C}}_{d}=\textrm{diag}(c^{0}, c^{2}, c^{4}, c^{1},
c^{3}, c^{1}, c^{3}, c^{2}, c^{2})$ is a diagonal matrix and
$\mathbf{M}_{0}$ is given by \cite{Lallemand}
\begin{equation}\label{eq15}
\mathbf{M}_{0} = \left( {\begin{array}{*{20}{c}}
   1 & 1 & 1 & 1 & 1 & 1 & 1 & 1 & 1  \\
   { - 4} & { - 1} & { - 1} & { - 1} & { - 1} & 2 & 2 & 2 & 2  \\
   4 & { - 2} & { - 2} & { - 2} & { - 2} & 1 & 1 & 1 & 1  \\
   0 & 1 & 0 & { - 1} & 0 & 1 & { - 1} & { - 1} & 1  \\
   0 & { - 2} & 0 & 2 & 0 & 1 & { - 1} & { - 1} & 1  \\
   0 & 0 & 1 & 0 & { - 1} & 1 & 1 & { - 1} & { - 1}  \\
   0 & 0 & { - 2} & 0 & 2 & 1 & 1 & { - 1} & { - 1}  \\
   0 & 1 & { - 1} & 1 & { - 1} & 0 & 0 & 0 & 0  \\
   0 & 0 & 0 & 0 & 0 & 1 & { - 1} & 1 & { - 1}  \\
\end{array}} \right).
\end{equation}
Consequently, one can also obtain $\mathbf{m}^{(eq)}$ from
Eq.~(\ref{eq8}),
\begin{equation}\label{eq16}
\mathbf{m}^{({eq})} = (\phi, -4\phi c^{2}+3C_{2}, 3\phi
c^{4}-3C_{2}c^{2}, B_{x}, -B_{x}c^{2}, B_{y}, -B_{y}c^{2},
\bar{C}_{xx}-\bar{C}_{yy}, \bar{C}_{xy})^{\top},
\end{equation}
where $C_{2}=\bar{C}_{xx}+\bar{C}_{yy}$,
$\bar{C}_{\alpha\beta}=\sum\limits_{k}c_{k\alpha}c_{k\beta}\phi_{k}^{(eq)}=C_{\alpha\beta}+dc_{s}^{2}D_{\alpha\beta}(\phi)$. For this lattice model, the relaxation matrix $\mathbf{S}$ can be defined as
\begin{equation}\label{eq17}
\mathbf{S}=\left( {\begin{array}{*{20}{c}}
   s_{0} & 0 & 0 & 0 & 0 & 0 & 0 & 0 & 0  \\
   0 & s_{1} & 0 & 0 & 0 & 0 & 0 & 0 & 0  \\
   0 & 0 & s_{2} & 0 & 0 & 0 & 0 & 0 & 0  \\
   0 & 0 & 0 & s_{3} & 0 & s_{35} & 0 & 0 & 0  \\
   0 & 0 & 0 & 0 & s_{4} & 0 & 0 & 0 & 0  \\
   0 & 0 & 0 & s_{53} & 0 & s_{5} & 0 & 0 & 0  \\
   0 & 0 & 0 & 0 & 0 & 0 & s_{6} & 0 & 0  \\
   0 & 0 & 0 & 0 & 0 & 0 & 0 & s_{7} & 0  \\
   0 & 0 & 0 & 0 & 0 & 0 & 0 & 0 & s_{8}  \\
\end{array}} \right),
\end{equation}
where the diagonal element $s_{k}$ is the relaxation parameter corresponding to $k$th
moment $m_{k}=\sum\limits_{j}\mathbf{M}_{kj}\phi_{j}$, and the off-diagonal components ($s_{35}$ and $s_{53}$) correspond to the rotation of the principal axis of anisotropic diffusion \cite{Yoshida2010}.
Besides, based on Eq.~(\ref{eq14}), one can
easily obtain the following equation,
\begin{equation}\label{eq18}
\mathbf{M}^{-1}\mathbf{S M}=\mathbf{M}_{0}^{-1}\mathbf{S M}_{0},
\end{equation}
and simultaneously, the evolution Eq.~(\ref{eq2}) can be rewritten
as
\begin{eqnarray}\label{eq19}
\phi_{k}(\mathbf{x} + \mathbf{c}_{k}\delta t,\;t + \delta t) & = &  \phi_{k}(\mathbf{x},\;t )-(\mathbf{M}_{0}^{-1}\mathbf{S M}_{0})_{kj}[\phi_{j}(\mathbf{x},\;t) - \phi_{j}^{(eq)}(\mathbf{x},\;t)] \nonumber \\
& + & (\delta t+\frac{\delta t^{2}}{2}\bar{D}_{k})R_{k}(\mathbf{x}, t).
\end{eqnarray}

Based on Eq.~(\ref{eq13}b), we can rewrite the first-order equations
in $\epsilon$, but here we present the first, fourth, and sixth ones
since only these three equations are useful in the following process of
obtaining NACDE,
\begin{subequations}\label{eq20}
\begin{equation}
\partial_{t_1}\phi+\partial_{1x}B_{x}+\partial_{1y}B_{y}= R^{(1)},
\end{equation}
\begin{equation}
\partial_{t_1}B_{x}+\partial_{1x}\bar{C}_{xx}+\partial_{1y}\bar{C}_{xy}=\tilde{B}_{x}R^{(1)}-\frac{s_{3}m_{3}^{(1)}+s_{35}m_{5}^{(1)}}{\delta t},
\end{equation}
\begin{equation}
\partial_{t_1}B_{y}+\partial_{1x}\bar{C}_{xy}+\partial_{1y}\bar{C}_{yy}=\tilde{B}_{y}R^{(1)}-\frac{s_{53}m_{3}^{(1)}+s_{5}m_{5}^{(1)}}{\delta t}.
\end{equation}
\end{subequations}
If we introduce a matrix $\mathbf{A}$ and a vector $\mathbf{m}_{35}^{(1)}$, which are defined as
\begin{equation}\label{eq21}
\mathbf{A}=\left( {\begin{array}{*{20}{c}}
   s_{3} & s_{35}   \\
   s_{53} & s_{5}
\end{array}} \right),\ \ \  \
\mathbf{m}_{35}^{(1)}=\left( {\begin{array}{*{20}{c}}
   m_{3}^{(1)}  \\
   m_{5}^{(1)}
\end{array}} \right),
\end{equation}
then Eqs.~(\ref{eq20}b) and~(\ref{eq20}c) can be rewritten in a vector form,
\begin{equation}\label{eq22}
\partial_{t_1}\mathbf{B}+\nabla_{1}\cdot\bar{\mathbf{C}}=\tilde{\mathbf{B}}R^{(1)}-\frac{1}{\delta t}\mathbf{A}\mathbf{m}_{35}^{(1)}.
\end{equation}
Similarly, we can also use Eq.~(\ref{eq13}c) to derive the second-order equations in $\epsilon$, but only the first one corresponding to the conservative variable $\phi$ is presented,
\begin{equation}\label{eq23}
\partial_{t_2}\phi + \nabla_{1}\cdot[(\mathbf{I}-\frac{\mathbf{A}}{2})\mathbf{m}_{35}^{(1)}] = \frac{\delta t}{2}(\theta-1)\partial_{t_1}R^{(1)}
 + \frac{\delta t}{2}[(\gamma-1)\nabla_{1}\cdot(\tilde{\mathbf{B}}R^{(1)})].
\end{equation}
Substituting Eq.~(\ref{eq22}) into
Eq.~(\ref{eq23}), we have
\begin{eqnarray}\label{eq24}
\partial_{t_2}\phi & + & \delta t\nabla_{1}\cdot[-(\mathbf{I}-\frac{\mathbf{A}}{2})\mathbf{A}^{-1}(\partial_{t_1}\mathbf{B}+\nabla_{1}\cdot\bar{\mathbf{C}}-\tilde{\mathbf{B}}R^{(1)})]\nonumber \\
& = & \frac{\delta t}{2}(\theta-1)\partial_{t_1}R^{(1)}
 + \frac{\delta t}{2}[(\gamma-1)\nabla_{1}\cdot(\tilde{\mathbf{B}}R^{(1)})],
\end{eqnarray}
with the aid of Eqs.~(\ref{eq7}) and~(\ref{eq20}a), we can rewrite Eq.~(\ref{eq24})
as
\begin{eqnarray}\label{eq25}
\partial_{t_2}\phi & = & \nabla_{1}\cdot[(\mathbf{A}^{-1}-\frac{1}{2}\mathbf{I})dc_{s}^{2}\delta t\nabla_{1}\cdot\mathbf{D}]+\frac{\delta t}{2}(\theta-1)\partial_{t_1}R^{(1)}\nonumber \\
& + & \delta t\nabla_{1}\cdot\{[(\frac{\gamma}{2}\mathbf{I}-\mathbf{A}^{-1})\tilde{\mathbf{B}}-(\frac{1}{2}\mathbf{I}-\mathbf{A}^{-1})\mathbf{B}^{'}]R^{(1)}\},
\end{eqnarray}
where the diffusion tensor $\mathbf{K}$ is given by
\begin{equation}\label{eq26}
\mathbf{K}=d c_{s}^{2}(\mathbf{A}^{-1}-\frac{1}{2}\mathbf{I})\delta t.
\end{equation}
From above equation, it is clear that, for a fixed diffusion
tensor, the parameter $d$ can be used to adjust the relaxation
parameters.

Through combining the results at $t_{1}$ and $t_{2}$ scales, i.e.,  Eqs.~(\ref{eq20}a) and~(\ref{eq25}), we can recover the following NACDE,
\begin{eqnarray}\label{eq27}
\partial_{t} \phi & + & \nabla\cdot\mathbf{B}=\nabla\cdot[\mathbf{K}\cdot(\nabla\cdot\mathbf{D})]+R \nonumber \\
& + & \frac{\delta t}{2}(\theta-1)\partial_{t}R + \delta t\nabla\cdot\{[(\frac{\gamma}{2}\mathbf{I}-\mathbf{A}^{-1})\tilde{\mathbf{B}}-(\frac{1}{2}\mathbf{I}-\mathbf{A}^{-1})\mathbf{B}^{'}]R\}.
\end{eqnarray}
To derive correct NACDE, the parameter $\theta$ and the tensor $\tilde{\mathbf{B}}$ should be set as
\begin{equation}\label{eq28}
\theta=1, \ \ \tilde{\mathbf{B}}=(\frac{\gamma}{2}\mathbf{I}-\mathbf{A}^{-1})^{-1}(\frac{1}{2}\mathbf{I}-\mathbf{A}^{-1})\mathbf{B}^{'}.
\end{equation}
We noted that although above analysis is only carried out for the two-dimensional MRT model with D2Q9 lattice, it can be extended to three-dimensional model without any substantial difficulties.

Now we give some remarks on the present model.

\textbf{Remark I:} We first want to present some discussion on the diffusion tensor $\mathbf{K}$. Actually, if the diffusion tensor $\mathbf{K}$ is taken by $\mathbf{K}=\kappa\mathbf{I}$ with $\kappa$ being a constant or variable, the NACDE [Eq.~(\ref{eq1})] would be reduced to the NCDE considered in the previous work \cite{Shi2009}, and can still be solved in the framework of the BGK model. However, if $\mathbf{K}$ is a diagonal matrix or full matrix where the element $\kappa_{ij}$ is a function of space, the MRT model rather than BGK model should be adopted. We would also like to point out that for the special case where $\mathbf{K}$ is a diagonal matrix,
\begin{equation}\label{29}
\mathbf{K}=\left( {\begin{array}{*{20}{c}}
   \kappa_{xx} & 0   \\
   0 & \kappa_{yy}
\end{array}} \right),
\end{equation}
the relaxation matrix $\mathbf{S}$ and the matrix $\mathbf{A}$ are also diagonal matrices ($s_{35}=s_{53}=0$), then the tensor $\tilde\mathbf{B}$ and the relation between the nonzero elements of $\mathbf{K}$ and relaxation parameters can be written in simple forms,
\begin{subequations}\label{eq30}
\begin{equation}
\tilde\mathbf{B}=\left( {\begin{array}{*{20}{c}}
   \lambda_{1} & 0   \\
   0 & \lambda_{2}
\end{array}} \right)\mathbf{B}^{'}, \ \ \  \lambda_{1}=\frac{s_{3}-2}{\gamma s_{3}-2}, \lambda_{2}=\frac{s_{5}-2}{\gamma s_{5}-2},
\end{equation}
\begin{equation}
\kappa_{xx}=d c_{s}^{2}(\frac{1}{s_{3}}-\frac{1}{2})\delta t,\ \ \ \kappa_{yy}=d c_{s}^{2}(\frac{1}{s_{5}}-\frac{1}{2})\delta t.
\end{equation}
\end{subequations}

\textbf{Remark II:} Based on the choice of the parameter $\gamma$, two \emph{special} schemes of the present model can be obtained.

\noindent
\textbf{Scheme A} ($\gamma=0$): $\tilde{\mathbf{B}}=(\mathbf{I}-\frac{1}{2}\mathbf{A})\mathbf{B}^{'}$.
 For this special case, a time derivative is included in second term of the right hand side of the evolution equation [see Eq.~(\ref{eq2})], and thus we need to use the finite-difference method to compute the time derivative term $\partial_{t}R_{k}(\mathbf{x}, t)$. Here for simplicity an explicit finite-difference scheme, i.e., $\partial_{t}R_{k}(\mathbf{x}, t)=[R_{k}(\mathbf{x}, t)-R_{k}(\mathbf{x}, t-\delta t)]/\delta t$, is adopted. Although a little larger memory cost would be caused for this scheme, the collision process can still be conducted locally.

\noindent
\textbf{Scheme B} ($\gamma=1$): $\tilde{\mathbf{B}}=\mathbf{B}^{'}$.
Under the present choice of the parameter $\gamma$, $\bar{D}_{k}=D_{k}=\partial_{t} + \mathbf{c}_{k}\cdot\nabla$, both the time derivative and the space derivative are contained in the evolution equation. Although we can still use explicit finite-difference schemes to calculate time and the space derivatives $(\partial_{t}+\mathbf{c}_{k}\cdot\nabla)R_{k}$, which is similar to the approach used in \textbf{Scheme A}, the collision process cannot be performed locally. To solve the problem, an implicit finite-difference scheme can be applied to compute $D_{k}R_{k}(\mathbf{x}, t)$,
\begin{equation}\label{eq31}
D_{k}R_{k}(\mathbf{x}, t)=\frac{R_{k}(\mathbf{x}+\mathbf{c}_{k}\delta t, t+\delta t)-R_{k}(\mathbf{x}, t)}{\delta t},
\end{equation}
which not only can result in that the collision process can be implemented locally, but also cause the implementation of present model to be explicit.
Substituting Eq.~(\ref{eq31}) into the evolution equation, we can obtain
\begin{eqnarray}\label{eq32}
\phi_{k}(\mathbf{x}+\mathbf{c}_{k}\delta t, \;t + \delta t)&-&\phi_{k}(\mathbf{x},\;t )= -(\mathbf{M}^{-1}\mathbf{S M})_{kj}[\phi_{j}(\mathbf{x},\;t) - \phi_{j}^{(eq)}(\mathbf{x},\;t)]\nonumber \\
& + &\frac{\delta
t}{2}[R_{k}(\mathbf{x}+\mathbf{c}_{k}\delta t, t+\delta
t)+R_{k}(\mathbf{x}, t)].
\end{eqnarray}
To avoid the implicitness, a new variable is introduced \cite{He1998},
\begin{equation}\label{eq33}
\bar{\phi}_{k}=\phi_{k}-\delta t R_{k}/2,
\end{equation}
then one can rewrite evolution equation as
\begin{eqnarray}\label{eq34}
\bar{\phi}_{k}(\mathbf{x} + \mathbf{c}_{k}\delta t,\;t + \delta t) & = & \bar{\phi}_{k}(\mathbf{x},\;t ) -(\mathbf{M}^{-1}\mathbf{S M})_{kj}[\bar{\phi}_{j}(\mathbf{x},\;t) - \phi_{j}^{(eq)}(\mathbf{x},\;t)]\nonumber \\
& + &\delta t[\mathbf{M}^{-1}(\mathbf{I}-\frac{\mathbf{S}}{2})\mathbf{M}]_{kj}R_{j}(\mathbf{x}, t).
\end{eqnarray}
which is the same as the evolution appeared in Ref. \cite{Chai2014}.
Based on the Eqs.~(\ref{eq10}a) and~(\ref{eq33}), the variable
$\phi$ in \textbf{Scheme B} can be calculated by
\begin{equation}\label{eq35}
\phi=\sum\limits_k\phi_{k}  = \sum\limits_k\bar{\phi}_{k}+\frac{\delta t}{2}R.
\end{equation}
Here it should be noted that if the source term $R$ is a
function of $\phi$, usually one needs to use some other methods to
solve the algebraic equation~(\ref{eq35}).

\textbf{Remark III:} We noted that there is still a limitation in the applications of above MRT model since it may be difficult or impossible to derive the function $\mathbf{C}$ analytically [see Eq.~(\ref{eq7})]. Following the idea presented in the work \cite{Chopard}, however, one can solve the problem through adding a new source term $G_{k}$ in the evolution equation, i.e.,
\begin{eqnarray}\label{eq36}
\phi_{k}(\mathbf{x} + \mathbf{c}_{k}\delta t,\;t + \delta t) & = &  \phi_{k}(\mathbf{x},\;t )-(\mathbf{M}^{-1}\mathbf{S M})_{kj}[\phi_{j}(\mathbf{x},\;t) - \phi_{j}^{(eq)}(\mathbf{x},\;t)] \nonumber \\
& + & (\delta t+\frac{\delta t^{2}}{2}\bar{D}_{k})R_{k}(\mathbf{x},\;t )+\delta t
G_{k}(\mathbf{x},\;t ),
\end{eqnarray}
where $G_{k}$ is defined as
\begin{eqnarray}\label{eq37}
G_{k}=\omega_{k}\frac{\mathbf{c}_{k}\cdot (\mathbf{I}-\frac{1}{2}\mathbf{A})\partial_{t}\mathbf{B}}{c_{s}^{2}},
\end{eqnarray}
and meanwhile, the equilibrium distribution function $\phi_{k}^{(eq)}(\mathbf{x},\;t)$ can be simplified as
\begin{equation}\label{eq38}
\phi_{k}^{(eq)}(\mathbf{x},\;t)=\omega_{k}[\phi+\frac{\mathbf{c}_{k}\cdot \mathbf{B}}{c_{s}^{2}}+\frac{(d \mathbf{D}-\phi\mathbf{I}):(\mathbf{c}_{k}\mathbf{c}_{k}-c_{s}^{2}\mathbf{I})}{2c_{s}^{2}}],
\end{equation}
which can be derived from Eq.~(\ref{eq3}) through setting $\mathbf{C}=\mathbf{0}$.
Through the Chapman-Enskog analysis, one can also find that Eq.~(\ref{eq1}) can be recovered correctly from Eq.~(\ref{eq36}). In addition, we would also like to point out that if $\mathbf{D}=\phi\mathbf{I}$, the D$n$Q$(2n)$ and D$n$Q$(2n+1)$ lattice models can also be used.

\textbf{Remark VI:} Although there are many LB models for CDEs \cite{Dawson, Guo1999, Sman, He, Zhang, Deng, Suga, Zheng, Shi2008,
Chopard, Huang2011, Chai2013, Perko, Ginzburg2005a, Ginzburg2005b,
Ginzburg2012, Servan-Camas, Rasin, Yoshida2010, Yoshida2014, Li,
Huang2014, Chai2014, Liang, Huang2015, Shi2009, Shi2011, Zhang2012}, most of them are limited to the linear CDEs with isotropic diffusion
\cite{Dawson, Guo1999, Sman, He, Deng, Suga, Zheng, Shi2008,
Chopard, Huang2011, Chai2013, Perko,
Servan-Camas, Rasin, Yoshida2014, Li,
Huang2014, Chai2014, Liang, Huang2015}, and what is more, some of them cannot give correct CDE \cite{Dawson, He, Zhang, Suga, Perko, Ginzburg2005a, Zhang2012}. Actually in the past decade, some LB models for anisotropic CDEs have also been developed \cite{Rasin, Zhang, Ginzburg2005a, Yoshida2010, Li, Huang2014}, but usually they can only be used to solve anisotropic CDEs where the convection term $\mathbf{B}$ or diffusion term $\mathbf{D}$ is a linear function of $\phi$ \cite{Rasin, Zhang, Yoshida2010, Li, Huang2014}. We also note that the LB model proposed by Ginzburg \cite{Ginzburg2005a} can be used to solve the CDEs with nonlinear convection and diffusion terms, but some additional
assumptions have been adopted to recover the correct CDE, as pointed out in Ref. \cite{Shi2009}. Recently, Shi and Guo proposed a new BGK model for NCDE \cite{Shi2009}, but the model is usually used to solve the isotropic NCDE, and cannot be directly applied to solve the NACDE. From above discussion, however, it is clear that the present MRT model can be viewed as a general LB model for the NACDE.

\section{Numerical results and discussion}

To test the accuracy and stability of present MRT model for NACDEs,
some classic examples, including the isotropic convection-diffusion
equation with a constant velocity, Burgers-Fisher equation,
Buckley-Leverett equation, and
anisotropic convection-diffusion
equations, will be considered in this section. In our
simulations, the following global relative error ($GRE$) defined by
Eq.~(\ref{eq39}) is used to test the accuracy of the present MRT
model,
\begin{equation}\label{eq39}
GRE=\frac{\sum\limits_\mathbf{x}|\phi_{a}(\mathbf{x}, t)-\phi_{n}(\mathbf{x}, t)|}{\sum\limits_\mathbf{x}|\phi_{a}(\mathbf{x}, t)|},
\end{equation}
where the subscripts $a$ and $n$ denote the analytical
and numerical solutions. The distribution function $\phi_{k}$ is
initialized by its equilibrium distribution function $\phi_{k}^{(eq)}$, i.e.,
\begin{equation}\label{eq40}
\phi_{k}(\mathbf{x}, t = 0) = \phi_{k}^{(eq)}|_{t = 0}.
\end{equation}
Unless otherwise stated, the parameter $d$ appeared in the equilibrium
distribution function is set to be 1.0, the \textbf{Scheme B} is
adopted since the computation of the time derivative in
\textbf{Scheme A} can be avoided, and meanwhile, the non-equilibrium
extrapolation scheme \cite{Guo2002} is adopted since it can be used to treat different boundary conditions and
also has a second-order convergence rate in space. In addition, it
should be noted that, besides the relaxation parameters ($s_{3}$, $s_{35}$,
$s_{5}$ and $s_{53}$) related to diffusion tensor, the other relaxation
parameters are simply taken as \cite{Chai2014}
\begin{equation}\label{eq41}
s_{0}=0,\ s_{1}=s_{2}=s_{4}=s_{6}=s_{7}=s_{8}=1.0.
\end{equation}

\subsection{Isotropic convection-diffusion equation with a constant velocity}

We first consider a simple two-dimensional isotropic CDE with a constant velocity, which can be expressed as
\begin{equation}\label{eq42}
\partial_{t}\phi+u_{x}\partial_{x}\phi+u_{y}\partial_{y}\phi=\kappa(\partial_{xx}\phi+\partial_{yy}\phi)+R,
\end{equation}
where $u_{x}$ and $u_{y}$ are constants, and are set to be 0.1,
$\kappa$ is the diffusion coefficient. $R$ is the source term, and
is given by
\begin{equation}\label{eq43}
R=\exp[(1-2\pi^{2}\kappa)t]\{\sin[\pi(x+y)]+\pi(u_{x}+u_{y})\cos[\pi(x+y)]\}.
\end{equation}
Under the periodic boundary conditions adopted on the domain $[0, 2]\times[0, 2]$ and the following initial condition,
\begin{equation}\label{eq44}
\phi(x, y, t=0)=\sin[\pi(x+y)],
\end{equation}
we can derive the analytical solution of the problem,
\begin{equation}\label{eq45}
\phi(x, y, t)=\exp[(1-2\pi^{2}\kappa)t]\sin[\pi(x+y)].
\end{equation}
When the present MRT model is used to study this problem, the
function $\mathbf{B}$, $\mathbf{C}$, $\mathbf{D}$ and the diffusion tensor $\mathbf{K}$ should be given by $\mathbf{B}=\phi\mathbf{u}$ with $\mathbf{u}=(u_{x},
u_{y})^{\top}$, $\mathbf{C}=\phi\mathbf{u}\mathbf{u}$,
$\mathbf{D}=\phi\mathbf{I}$ and $\mathbf{K}=\kappa\mathbf{I}$.

We now performed some simulations under different time and different
P\'{e}clet numbers (Pe=$Lu_{x}/\kappa$, $L=2.0$ is characteristic
length), and presented the results in Fig. 1 where the lattice size
is $201\times201$. As seen from the figure, the numerical results
are in good agreement with analytical solutions.
\begin{figure}
\includegraphics[width=3in]{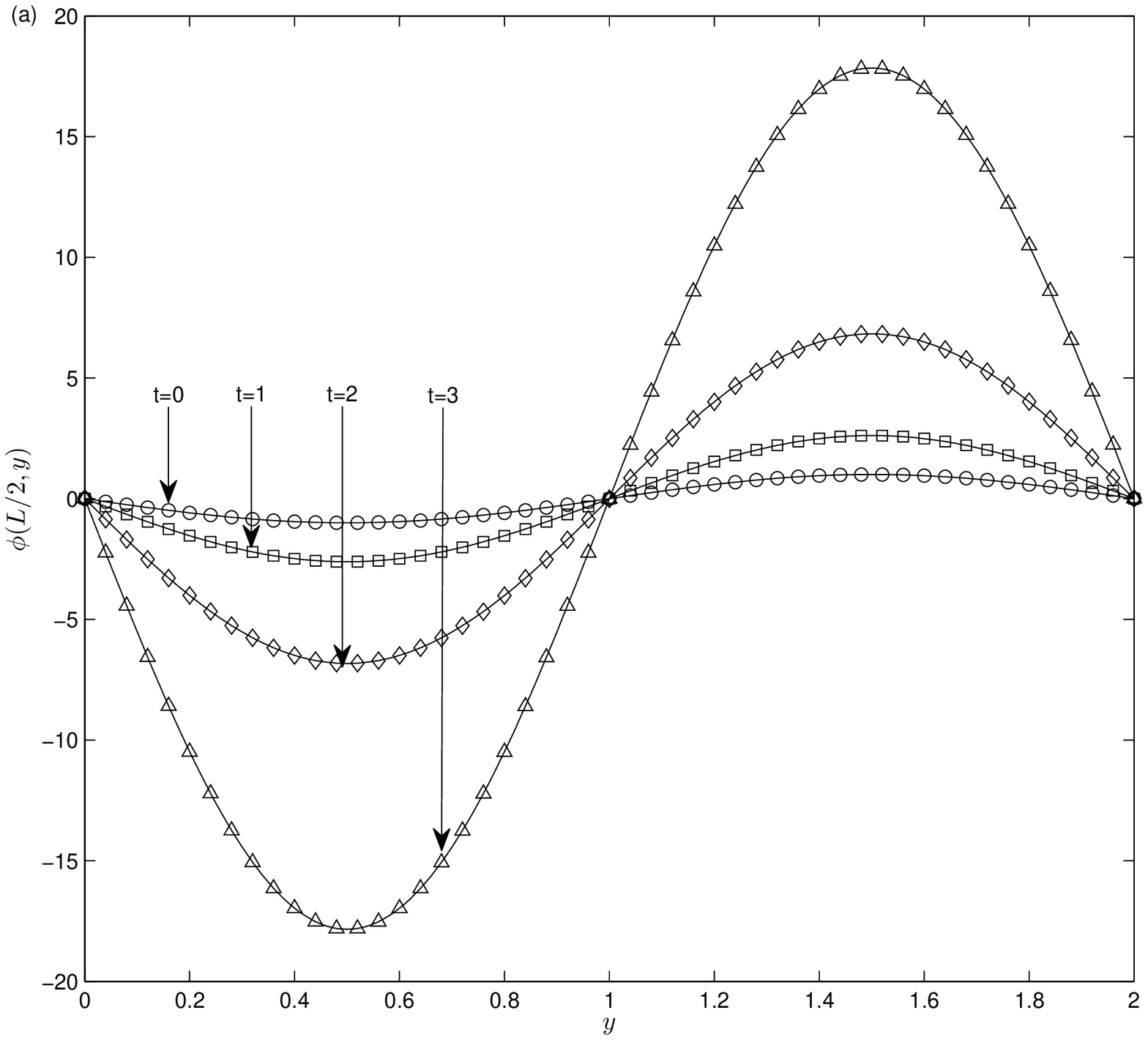}
\includegraphics[width=3in]{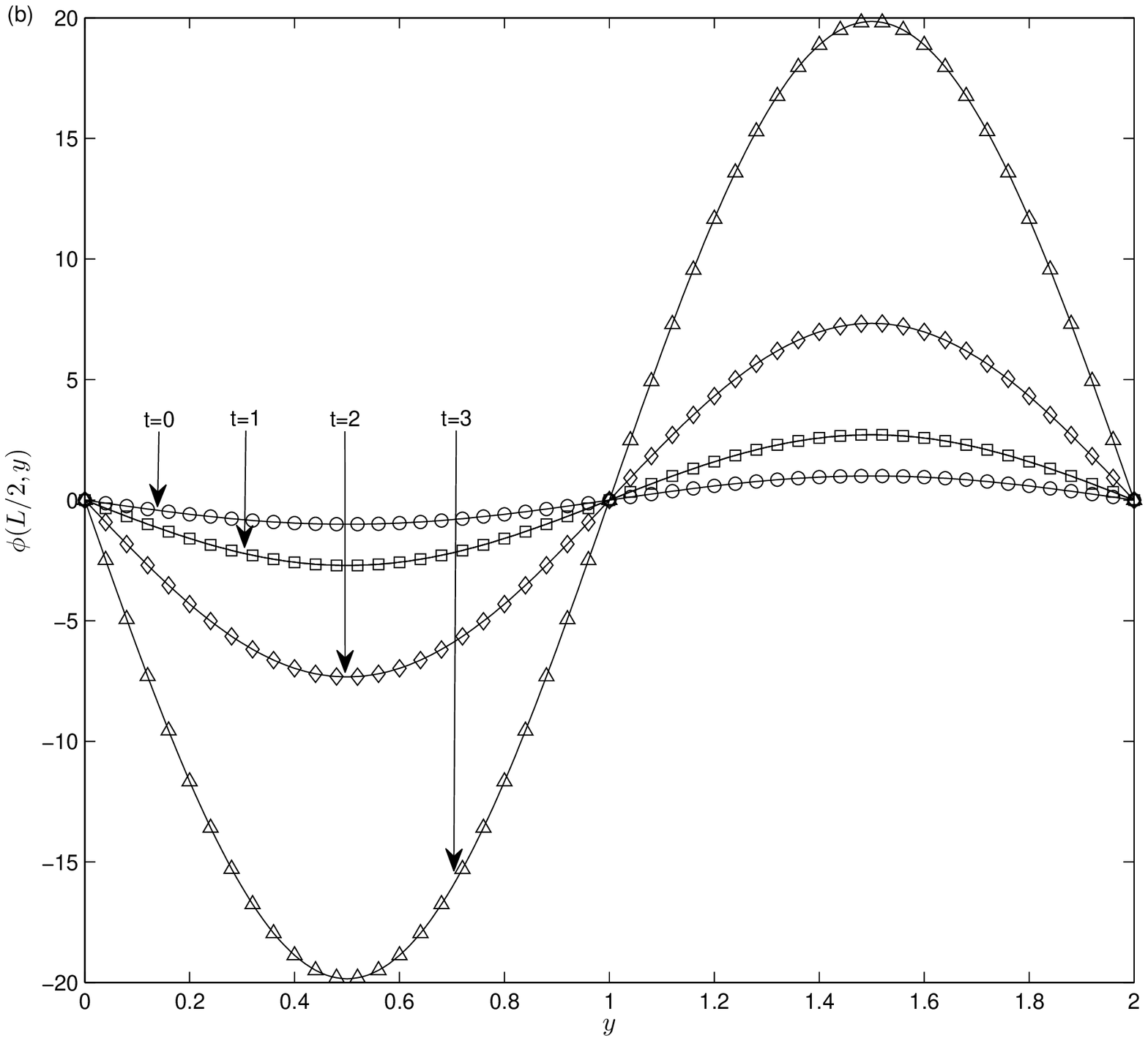}
\caption{\label{fig:1} Profiles of scalar variable $\phi$ at
different time and P\'{e}clet numbers [(a): Pe=100, (b): Pe=1000;
solid lines: analytical solutions, symbols: numerical results].}
\end{figure}
Besides, this problem is also applied to test the convergence rate of the
present model since the boundary effect can be eliminated by the periodic boundary conditions adopted. To this end, we
conducted a number of simulations and computed the $GRE$s under
different lattice resolutions. As shown in Fig. 2 where
lattice spacing is varied from $L/500$ to $L/100$ and numerical
simulations are suspended at time $T=3.0$, it is clear that present MRT
model has a second-order convergence rate in space.
\begin{figure}
\includegraphics[width=3in]{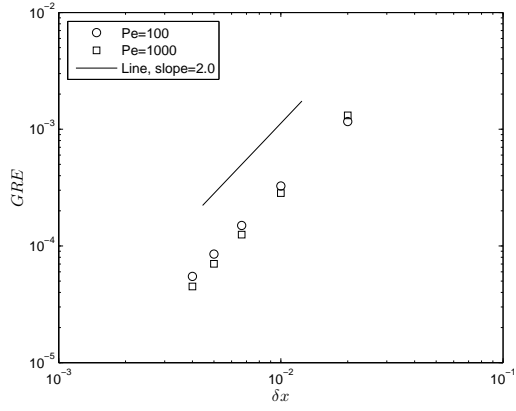}
\centering \caption{\label{fig:2} The global relative errors at
different lattice sizes ($\delta x=L/500$, $L/400$, $L/300$, $L/200$
and $L/100$), the slope of the inserted line is 2.0, which indicates
the present MRT model has a second-order convergence rate in space.}
\end{figure}

\subsection{The two-dimensional Burgers-Fisher equation}

The two-dimensional Burgers-Fisher equation, as a special case of the
NACDEs, can be written as \cite{Shi2009, Wazwaz2005}
\begin{equation}\label{eq46}
\partial_{t}\phi+a\phi^{\delta}\partial_{x}\phi=\kappa(\partial_{xx}\phi+\partial_{yy}\phi)+R,
\end{equation}
where $R=b\phi(1-\phi^{\delta})$ is the source term, $\delta$, $a$,
$b$ and diffusion coefficient $\kappa$ are constants. Compared to
the first problem considered above, the present problem is more
complicated since it is nonperiodic and nonlinear, but we
can still obtain its analytical solution under the proper initial and boundary
conditions,
\begin{equation}\label{eq47}
\phi(x, y, t)=\{\frac{1}{2}+\frac{1}{2}\tanh[A(x+y-\omega
t)]\}^{\frac{1}{\delta}},
\end{equation}
where $A$ and $\omega$ are two parameters, and are defined by
\begin{equation}\label{eq48}
A=-\frac{a\delta}{4\kappa(\delta+1)}, \ \
\omega=\frac{a^{2}+2b\kappa(\delta+1)}{a(\delta +1)}.
\end{equation}

Compared with the NACDE defined by Eq.~(\ref{eq1}), the function
$\mathbf{B}(\phi)$ should be given by
\begin{equation}\label{eq49}
\mathbf{B}(\phi)=\phi^{\delta+1}(\frac{a}{\delta+1}, 0)^{\top}.
\end{equation}
Based on Eq.~(\ref{eq7}), one can further derive the tensor
$\mathbf{C}$ in Eq.~(\ref{eq3}),
\begin{equation}\label{eq50}
\mathbf{C}=\left( {\begin{array}{*{20}{c}}
   \frac{a^{2}}{2\delta+1}\phi^{2\delta+1} & 0  \\
   0 & 0 \\
\end{array}} \right).
\end{equation}
The diffusion tensor $\mathbf{K}$ and tensor $\mathbf{D}$ can be simply determined as
\begin{equation}\label{eq51}
\mathbf{K}=\kappa\mathbf{I},\ \ \mathbf{D}=\phi\mathbf{I}.
\end{equation}
We now consider how to use present MRT model to solve the
Burgers-Fisher equation. If the \textbf{scheme B} is adopted to
solve Eq.~(\ref{eq46}), one needs to use some other methods to
solve nonlinear equation (\ref{eq33}) since the source term $R$ is a
nonlinear function of $\phi$. To avoid such process, the
\textbf{Scheme A} is applied to solve the Burgers-Fisher equation.
We carried out some simulations in the computational domain $[-1,
2]\times[-1, 2]$, and presented numerical results and corresponding
analytical solutions under different time and different modified
P\'{e}clet numbers (Pe=$La/\kappa$, $L=3.0$ is the characteristic
length) in Fig.3 where the parameters $a$, $b$, and lattice size are
fixed to be 4.0, 1.0 and $301\times301$. As seen from the figure,
the numerical results agree well with corresponding analytical solutions.

\begin{figure}
\includegraphics[width=3in]{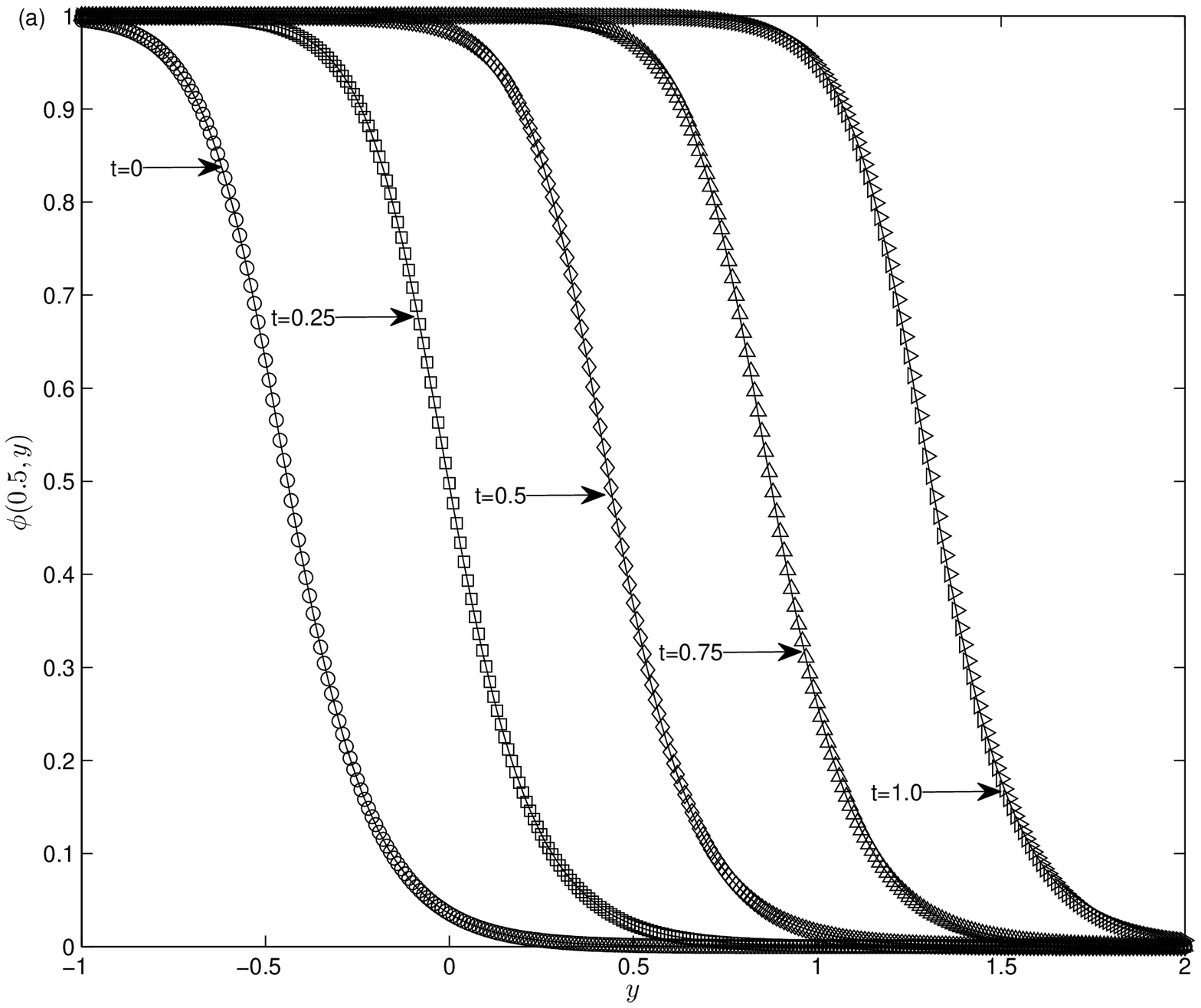}
\includegraphics[width=3in]{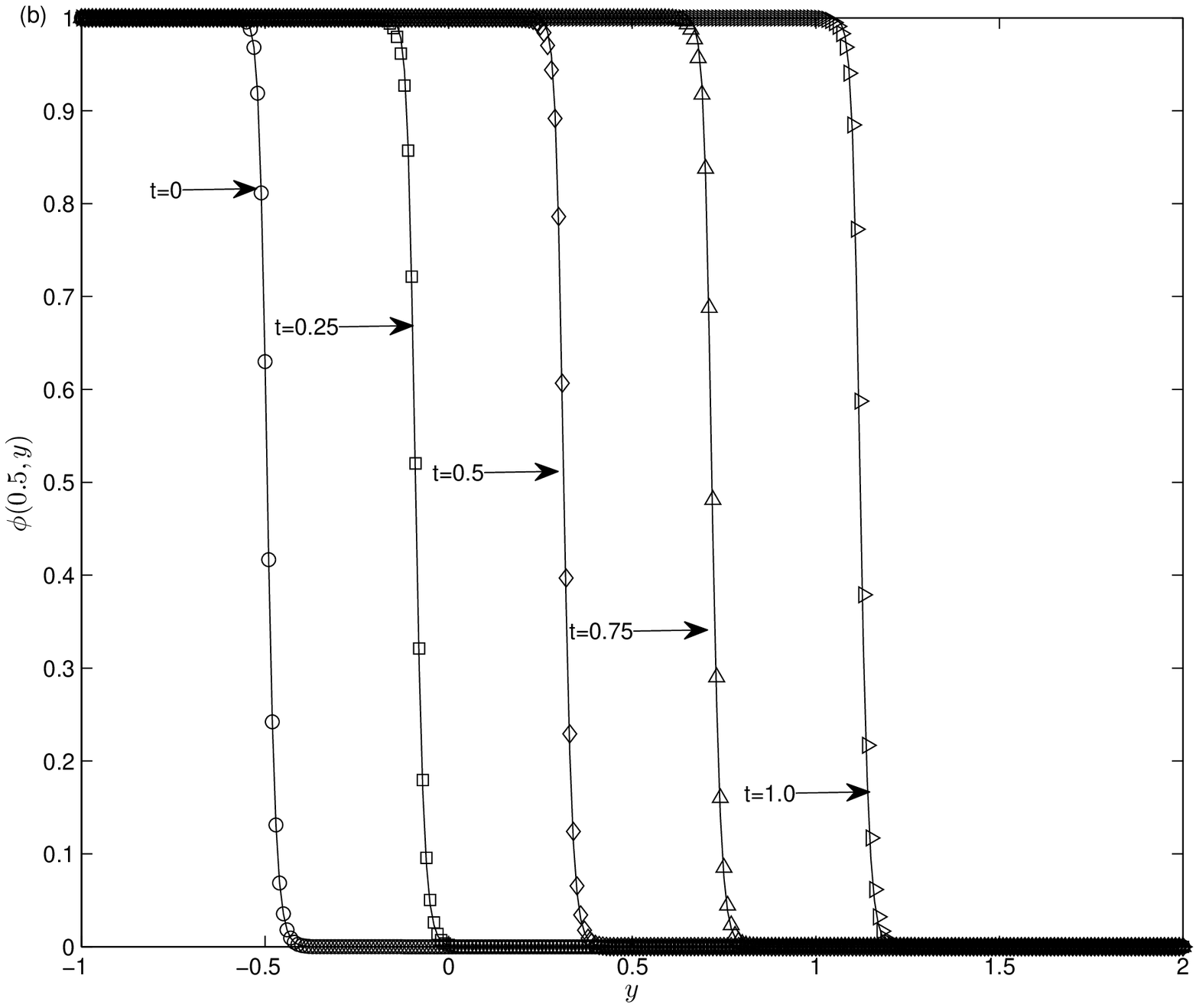}
\caption{\label{fig:3} Profiles of the scalar variable $\phi$ at
different time and modified
P\'{e}clet numbers [(a): Pe=100, (b):
Pe=1000; solid lines: analytical solutions, symbols: numerical
results].}
\end{figure}

We note that this problem is nonperiodic and the boundary
effect cannot be excluded, thus it can be used to test the
convergence rate of the MRT model coupling with the non-equilibrium
extrapolation scheme. To this end, we also carried out some simulations, and presented the $GRE$s in Fig. 4 where the lattice
spacing is ranged from $L/900$ to $L/300$ and simulations are
suspended at time $T=1.0$. As shown in this figure, the present MRT
model coupling with non-equilibrium extrapolation scheme also has a
second-order convergence rate in space. Besides, it is also found
that the $GREs$ at Pe=600 is less than those at Pe=120, this may be
because the relaxation parameters ($s_{3}$ and $s_{5}$) corresponding to the case of Pe=600 are
more close to 1, which usually give more accurate results.
\begin{figure}
\includegraphics[width=3in]{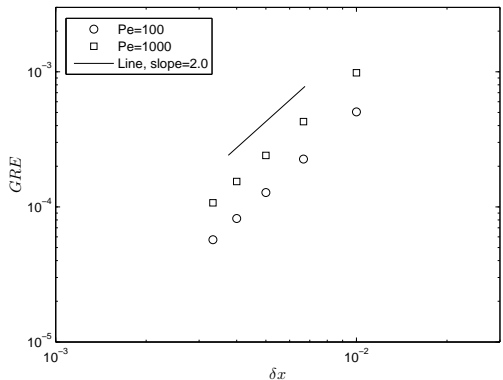}
\centering \caption{\label{fig:4} The global relative errors at
different lattice sizes ($\delta x=L/900$, $L/750$, $L/600$, $L/450$
and $L/300$), the slope of the inserted line is 2.0, indicating that
the present MRT model has a second-order convergence rate in space.}
\end{figure}

\subsection{The two-dimensional Buckley-Leverett equation}

We also consider the two-dimensional Buckley-Leverett equation
\cite{Shi2011, Karlsen1998, Kurganov2000}
\begin{equation}\label{eq52}
\partial_{t}\phi+\partial_{x}f(\phi)+\partial_{y}g(\phi)=\kappa(\partial_{xx}\phi+\partial_{yy}\phi),
\end{equation}
with the following initial condition,
\begin{equation}\label{eq53}
\phi(x, y, 0)=\left\{
\begin{array}{ll}
 1,                        & x^2+y^2<0.5,   \\
 0,                        & x^2+y^2\geq0.5,
\end{array}\nonumber\\
\right.
\end{equation}
where $\kappa$ is the diffusion coefficient. $f(\phi)$ and $g(\phi)$
are a function of $\phi$, and are defined as
\begin{equation}\label{eq54}
f(\phi)=\frac{\phi^{2}}{\phi^{2}+(1-\phi)^2},\ \ \
g(\phi)=f(\phi)[1-5(1-\phi)^2].
\end{equation}
We note that, similar to the Burgers-Fisher equation considered previously, the Buckley-Leverett equation is also a special
NACDE, but there is no analytical solution to this problem.

When the present MRT model is used to solve the problem, the evolution equation [Eq.~(\ref{eq2})] can be written more simply since there is no source term included in the the Buckley-Leverett equation,
\begin{equation}\label{eq55}
\phi_{k}(\mathbf{x} + \mathbf{c}_{k}\delta t,\;t + \delta t) = \phi_{k}(\mathbf{x},\;t )-(\mathbf{M}^{-1}\mathbf{S M})_{kj}[\phi_{j}(\mathbf{x},\;t) - \phi_{j}^{(eq)}(\mathbf{x},\;t)],
\end{equation}
then from Eqs.~(\ref{eq1}) and~(\ref{eq7}), one can further determine
the functions $\mathbf{B}$, $\mathbf{C}$, $\mathbf{D}$ and diffusion tensor $\mathbf{K}$,
\begin{equation}\label{eq56}
\mathbf{B}=[f(\phi), g(\phi)]^{\top},\ \  \mathbf{C}=\left(
{\begin{array}{*{20}{c}}
   C_{xx} & C_{xy}  \\
   C_{yx} & C_{yy} \\
\end{array}} \right), \ \ \mathbf{D}=\phi\mathbf{I},\ \ \mathbf{K}=\kappa\mathbf{I},
\end{equation}
where the elements of the matrix $\mathbf{C}$ are given as
\begin{subequations}\label{eq57}
\begin{equation}
C_{xx}=\frac{1}{2}\arctan(2\phi-1)+\frac{1}{6A^3}(2-A+3A^2)(\phi-\frac{1}{2}),
\end{equation}
\begin{eqnarray}
C_{xy}=C_{yx} & = & \frac{1}{2}\arctan(2\phi-1)+\frac{5}{4}\ln(\frac{A}{2})\nonumber \\ & + & \frac{1}{2A}(2+\phi)+\frac{1}{12A^2}(\frac{17-4\phi}{2}+\frac{-7+4\phi}{A}),
\end{eqnarray}
\begin{eqnarray}
C_{yy} & = & -\frac{17}{8}\arctan(2\phi-1)+\frac{5}{2}\ln(\frac{A}{2})+25\phi[\frac{1}{4}-\frac{\phi}{2}+\frac{\phi^2}{3}]\nonumber \\ & + & \frac{1}{16A}(-89+258\phi)+\frac{1}{16A^2}(13+14\phi)+\frac{1}{24A^3}(1-42\phi).
\end{eqnarray}
\end{subequations}

In our simulations, the computational domain of the problem and the
lattice size are set to be $[-1.5, 1.5]\times[-1.5, 1.5]$ and
$301\times301$. We performed simulations at two different diffusion coefficients ($\kappa=0.1$ and $0.01$) or equivalently two different P\'{e}clet numbers (Pe=$LU/\kappa$=30 and 300, $L=3.0$ is the characteristic
length, $U=1.0$ is the characteristic velocity), and presented the results at time $T=0.5$ in Figs. 5 and 6. As expected, when the Pe is smaller or the diffusion coefficient is larger, the role of the diffusion becomes dominant, and thus the distribution of scalar variable $\phi$ is more smooth [see Fig. 6(a)]. Besides, to show more details on the distributions of scalar variable $\phi$ at different time, we also conducted some simulations, and presented the results of $\phi$ along the vertical centreline in Fig. 7. As seen from the figure, with the increase of time, the distribution of the scalar variable with a small Pe becomes more smooth.
\begin{figure}
\includegraphics[width=3in]{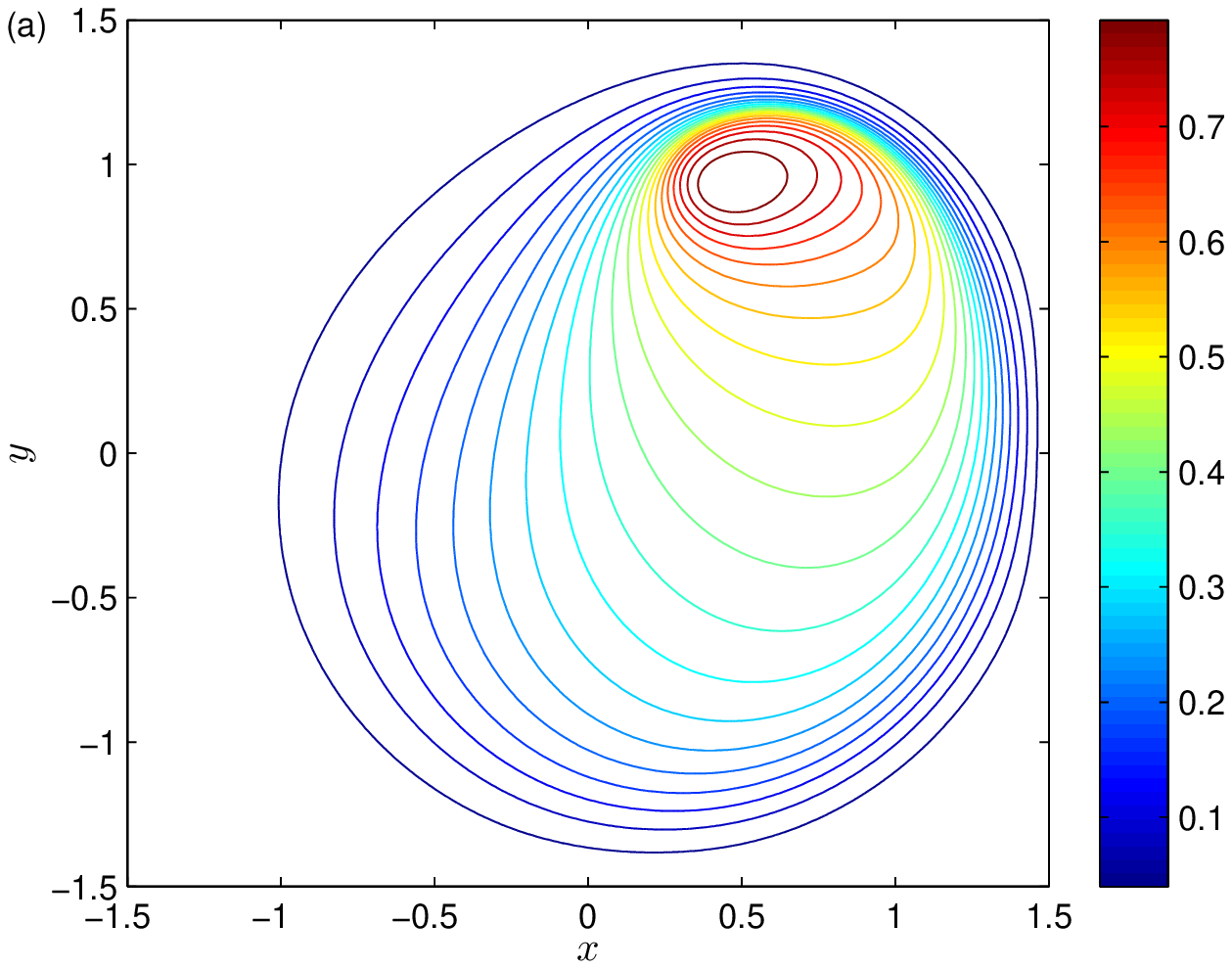}
\includegraphics[width=3in]{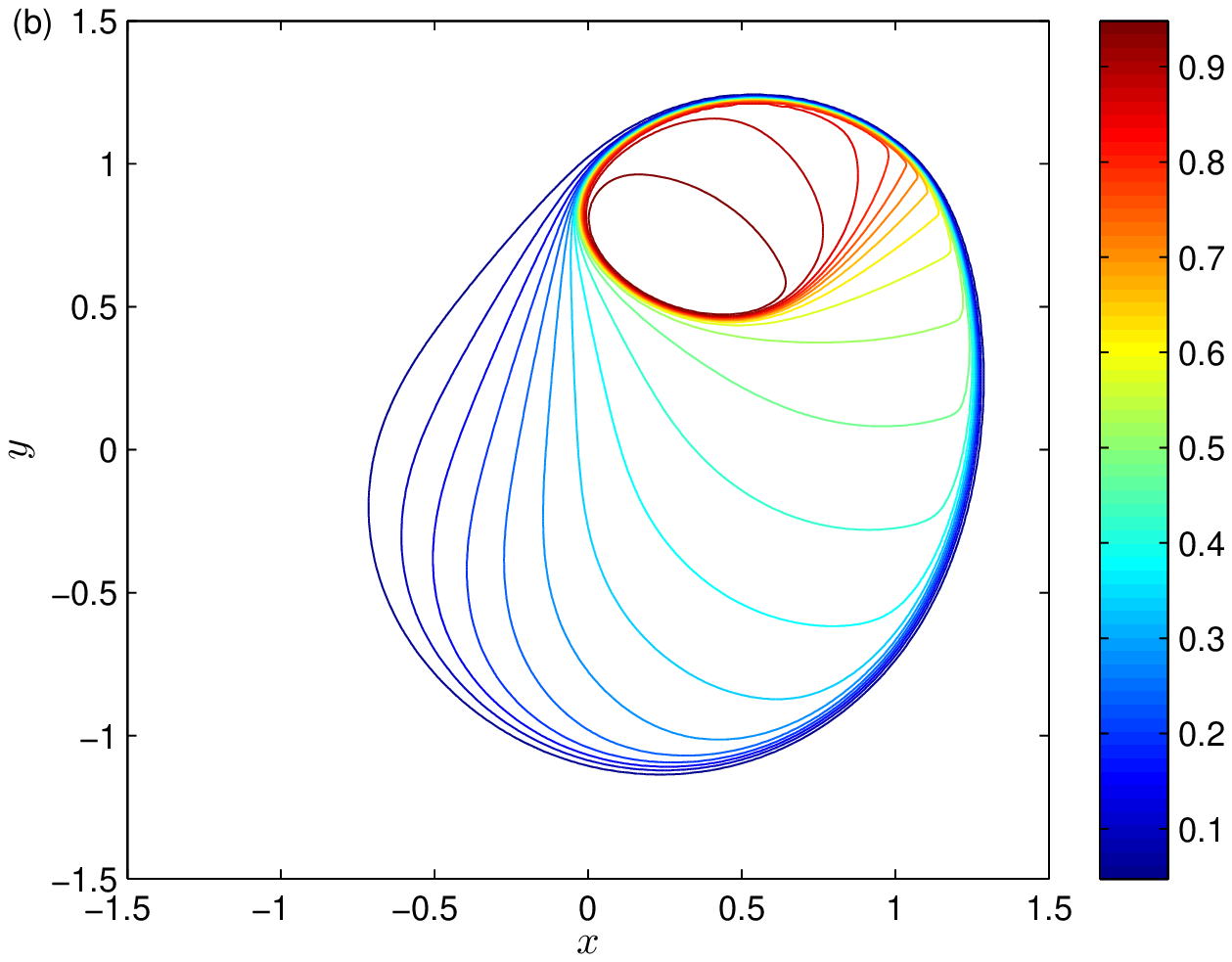}
\caption{\label{fig:5} Contours of the scalar variable $\phi$ at
the time $T=0.5$ and different P\'{e}clet numbers [(a): Pe=30, (b):
Pe=300].}
\end{figure}

\begin{figure}
\includegraphics[width=3in]{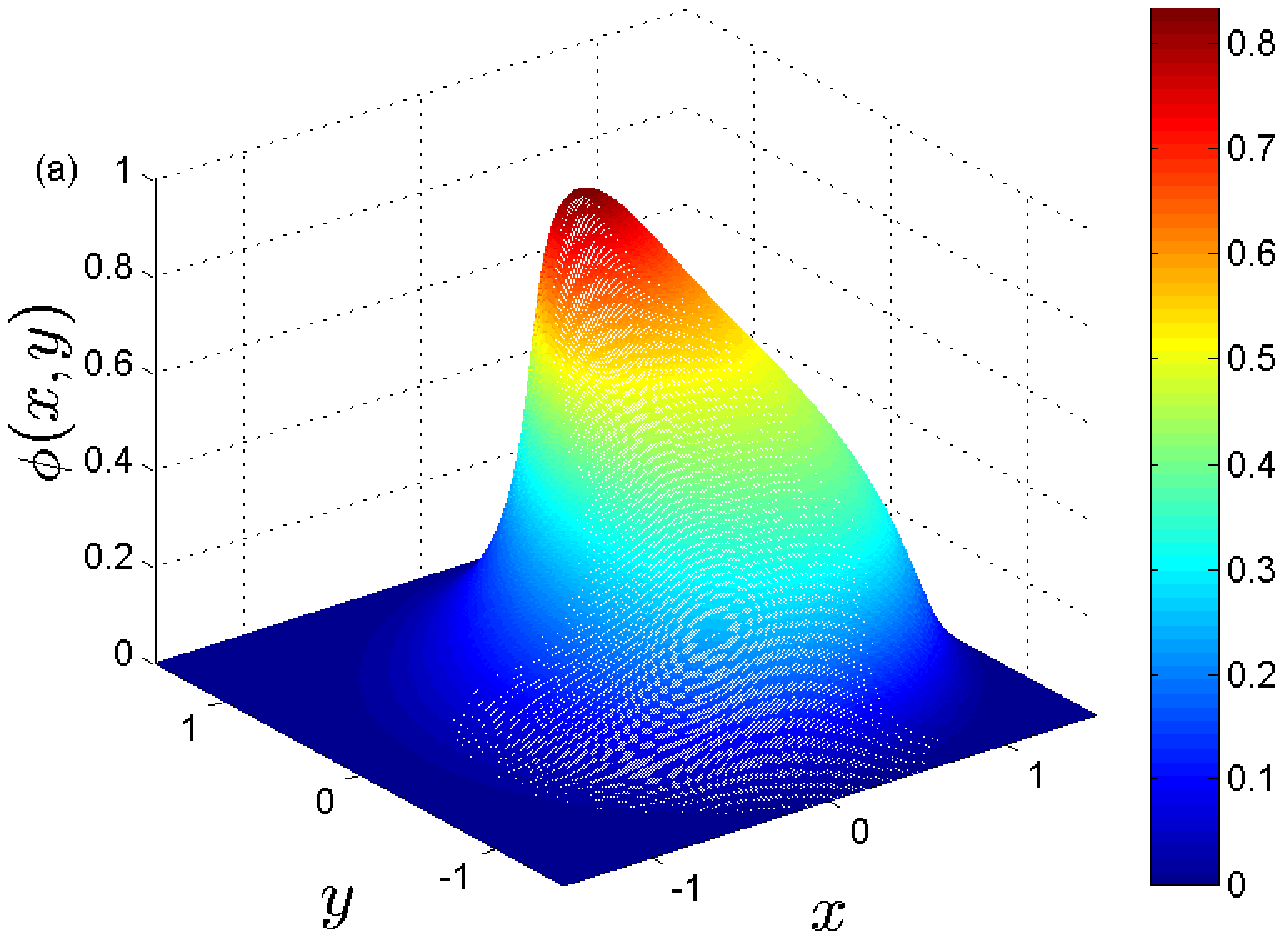}
\includegraphics[width=3in]{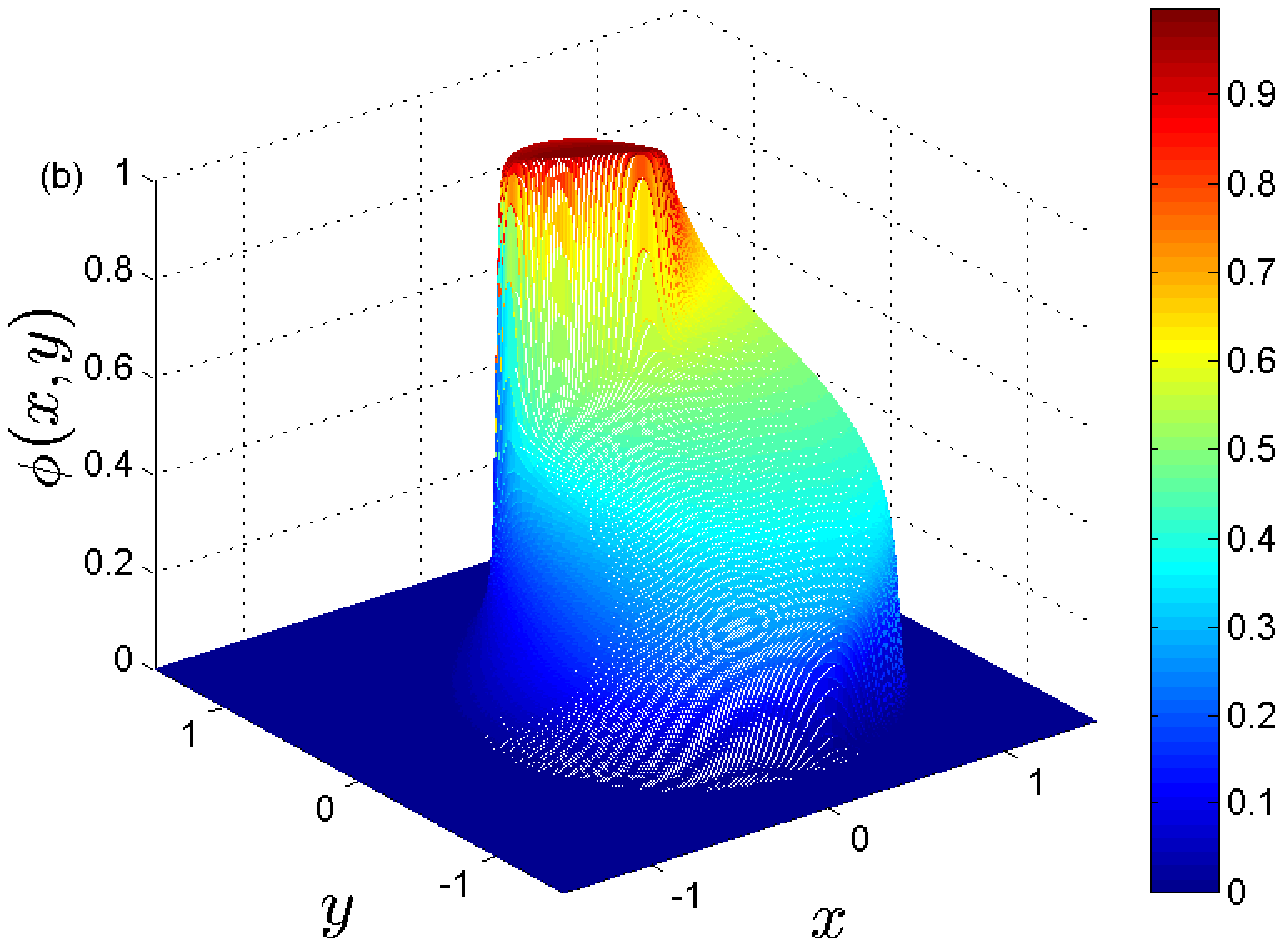}
\caption{\label{fig:6} Distributions of the scalar variable $\phi$ at
the time $T=0.5$ and different P\'{e}clet numbers [(a): Pe=30, (b):
Pe=300].}
\end{figure}

\begin{figure}
\includegraphics[width=3in]{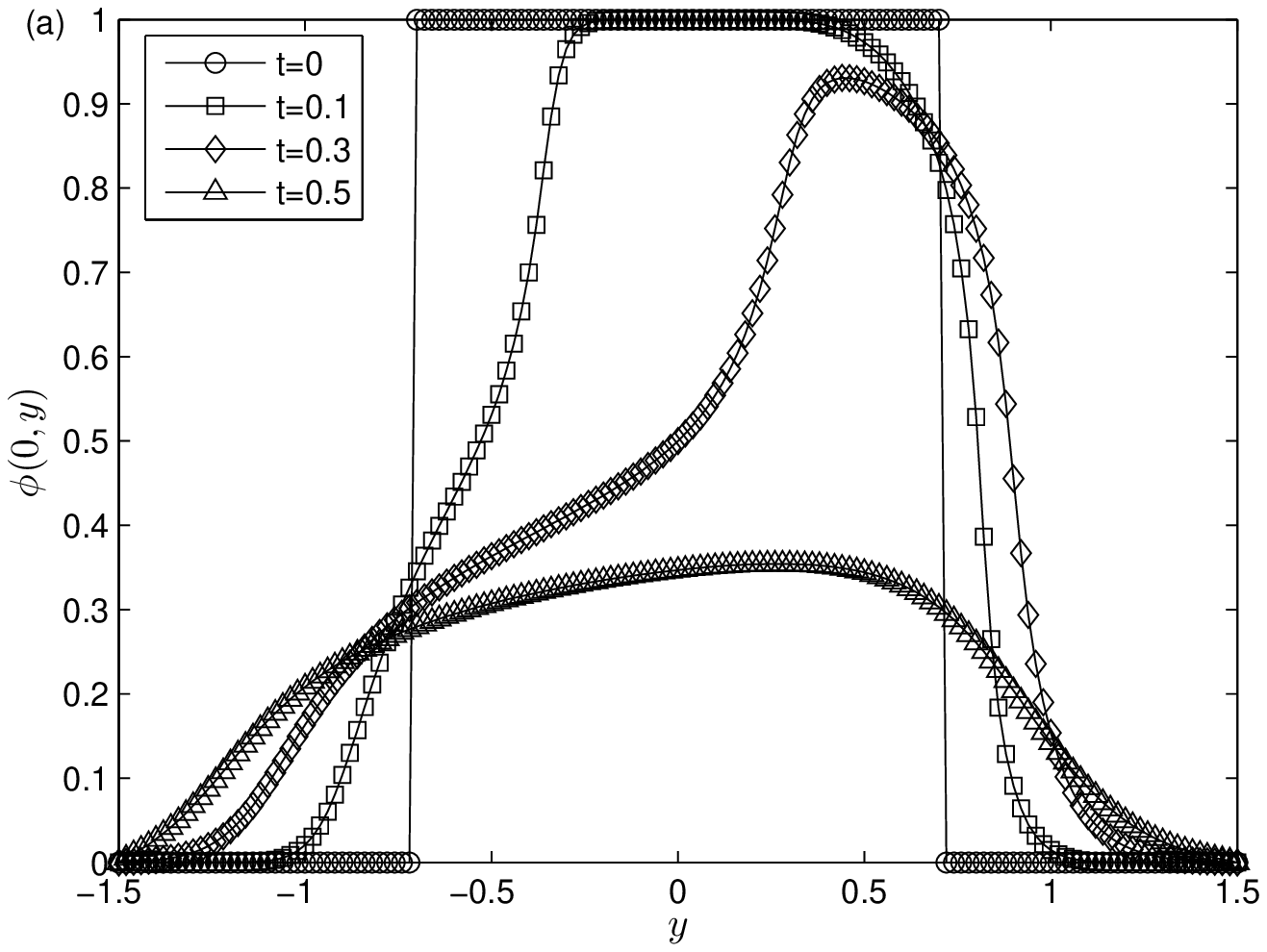}
\includegraphics[width=3in]{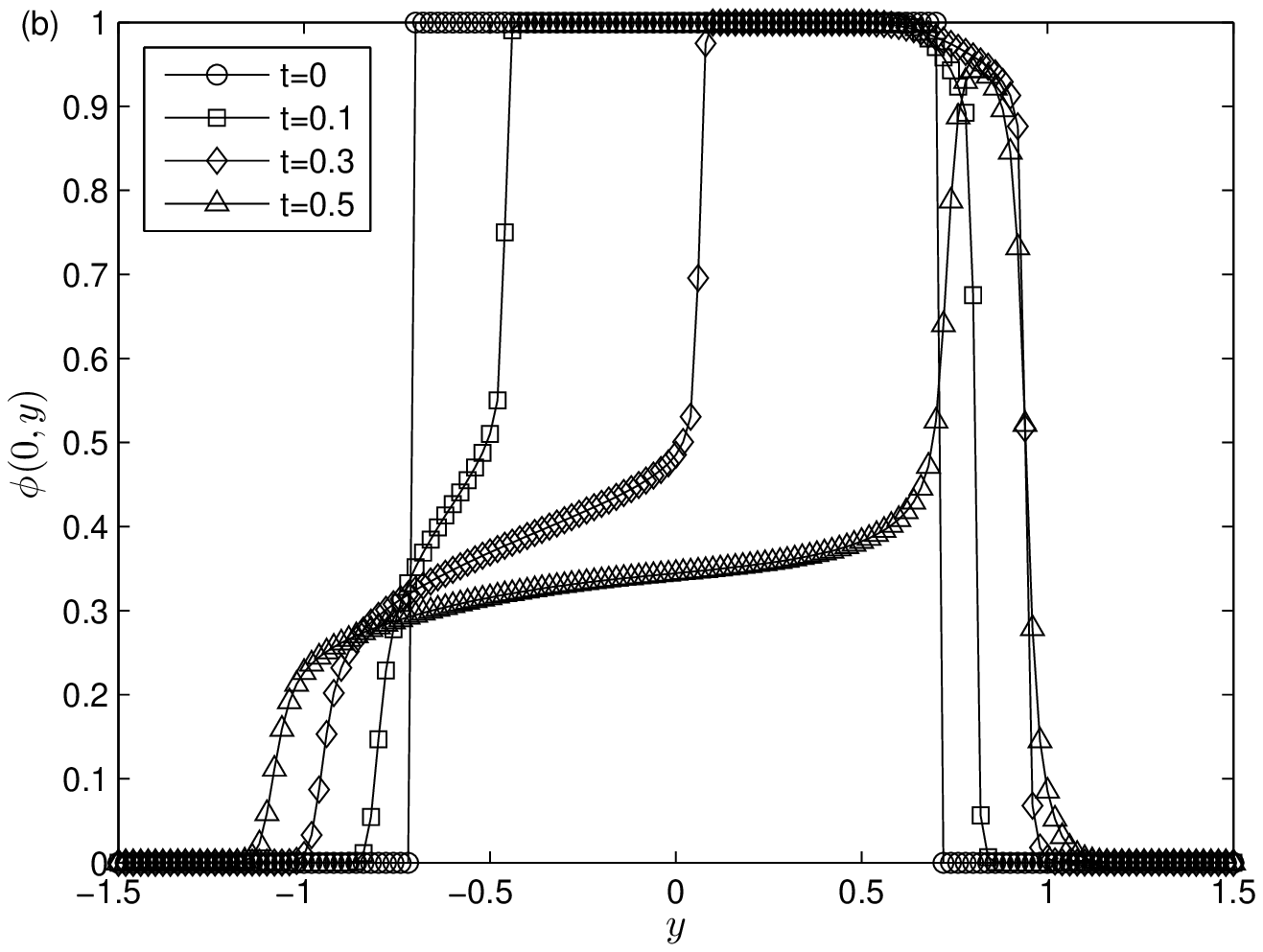}
\caption{\label{fig:7} Profiles of the scalar variable $\phi$ at
different time and P\'{e}clet numbers [(a): Pe=30, (b):
Pe=300].}
\end{figure}

We note that although the problem has no analytical solution, the present results
[see Figs. 5(b) and 6(b) where $\kappa=0.01$] agree well with those reported in some previous studies
\cite{Shi2011, Karlsen1998, Kurganov2000}, which can also be used to conclude that the present MRT model is also accurate in solving the Buckley-Leverett equation.

\subsection{Anisotropic convection-diffusion
equation with constant velocity and diffusion tensor}

We now consider the problem of the Gaussian hill with constant velocity and diffusion tensor, which is also a classic benchmark example and has also been used to validate LB models for anisotropic CDEs \cite{Ginzburg2005a, Servan-Camas, Yoshida2010, Huang2014}. The CDE for this problem can be written as
\begin{equation}\label{eq58}
\partial_{t}\phi+\nabla\cdot(\phi\mathbf{u})=\nabla\cdot(\mathbf{K}\cdot\nabla\phi),
\end{equation}
where $\mathbf{u}=(u_{x}, u_{y})^{\top}$ is a constant velocity, $\mathbf{K}$ is the constant diffusion tensor, and can be defined as
\begin{equation}\label{eq59}
\mathbf{K}=\left(
{\begin{array}{*{20}{c}}
   \kappa_{xx} & \kappa_{xy}  \\
   \kappa_{yx} & \kappa_{yy} \\
\end{array}} \right).
\end{equation}
Under the proper initial and boundary conditions, one can also derive the analytical solution of the problem,
\begin{equation}\label{eq60}
\phi(x, y, t)=\frac{\phi_{0}}{2\pi|\det(\sigma)|^{1/2}}\exp\{-\frac{\sigma^{-1}:[(\mathbf{x}-\mathbf{u}t)(\mathbf{x}-\mathbf{u}t)]}{2}\},
\end{equation}
where $\mathbf{x}=(x, y)^{\top}$, $\sigma=\sigma_{0}^{2}\mathbf{I}+2\mathbf{K}t$, $\sigma^{-1}$ is inverse matrix of $\sigma$, $|\det(\sigma)|$ is the absolute value of the determinant of $\sigma$.

Actually, there are two approaches that can be adopted to study the Gaussian hill problem. The first is that we directly use the MRT model to solve Eq.~(\ref{eq58}) with an anisotropic form, and set the functions $\mathbf{B}$, $\mathbf{C}$, and $\mathbf{D}$ as
\begin{equation}\label{eq61}
\mathbf{B}=(\phi u_{x}, \phi u_{y})^{\top},\ \  \mathbf{C}=\left(
{\begin{array}{*{20}{c}}
   \phi u_{x}^{2} & \phi u_{x}u_{y}  \\
   \phi u_{x}u_{y} & \phi u_{y}^{2} \\
\end{array}} \right), \ \ \mathbf{D}=\phi\mathbf{I}.
\end{equation}
While in the second approach, we first need to write Eq.~(\ref{eq58}) in an isotropic form,
\begin{equation}\label{eq62}
\partial_{t}\phi+\nabla\cdot(\phi\mathbf{u})=\nabla\cdot[\kappa(\nabla\cdot\mathbf{D})].
\end{equation}
which is then solved by the MRT model. In addition, it should be noted that Eq.~(\ref{eq62}) can also be solved by the previous BGK model \cite{Shi2009}. Based on Eq.~(\ref{eq62}), one can also find that the functions $\mathbf{B}$ and $\mathbf{C}$ should be the same as those appeared in Eq.~(\ref{eq61}), but the tensor $\mathbf{D}$ should be given by $\mathbf{D}=\mathbf{K}\phi/\kappa$ with $\kappa$ being a positive constant.

Similar to some previous works \cite{Yoshida2010, Huang2014}, we also considered the Gaussian hill problem in a bounded domain $[-1, 1]\times[-1, 1]$, and adopted the periodic boundary condition on all boundaries. In our simulations, $\sigma_{0}=0.01$ which is small enough to ensure that the periodic boundary condition adopted is reasonable and accurate at a finite time $T$, $\phi_{0}=2\pi\sigma_{0}^{2}$, $u_{x}=u_{y}=0.01$, and the lattice size is $401\times401$. To test the capacity of the present MRT model in solving the anisotropic CDEs, the following three types of diffusion tensor are considered,
\begin{equation}\label{eq63}
\mathbf{K}=\left[\left(
{\begin{array}{*{20}{c}}
   1 & 0  \\
   0 & 1 \\
\end{array}} \right),\ \
\left(
{\begin{array}{*{20}{c}}
   1 & 0  \\
   0 & 2 \\
\end{array}} \right), \ \
\left(
{\begin{array}{*{20}{c}}
   1 & 1  \\
   1 & 2 \\
\end{array}} \right)\right]\times10^{-3},
\end{equation}
which are usually denoted as isotropic, diagonally anisotropic and fully anisotropic diffusion problems.

We conducted some simulations by using above two approaches, and presented numerical results in Figs. 8, 9 and 10 where $\kappa=10^{-3}$ is used in the second approach. As seen from these figures, the numerical results qualitatively agree with analytical solutions. To quantitatively measure the deviations between numerical results and analytical solutions, the $GRE$s of isotropic, diagonally anisotropic and fully anisotropic diffusion problems are computed, and the values of them are $1.199\times10^{-4}$, $3.853\times10^{-4}$ and $6.531\times10^{-4}$ for the first approach, while they are $1.199\times10^{-4}$, $2.118\times10^{-4}$ and $4.572\times10^{-4}$ for the second approach, which illustrate that the present MRT model is accurate in studying these problems. Besides, the convergence rate of the MRT model for anisotropic CDEs is also investigated, and the results are shown in Fig. 11 where the lattice size is varied from $201\times201$ to $801\times801$. From this figure, one can find that, similar to some available MRT models for anisotropic diffusion problems \cite{Yoshida2010, Huang2014}, the present MRT model also has a second-order convergence rate in space.
\begin{figure}
\includegraphics[width=3in]{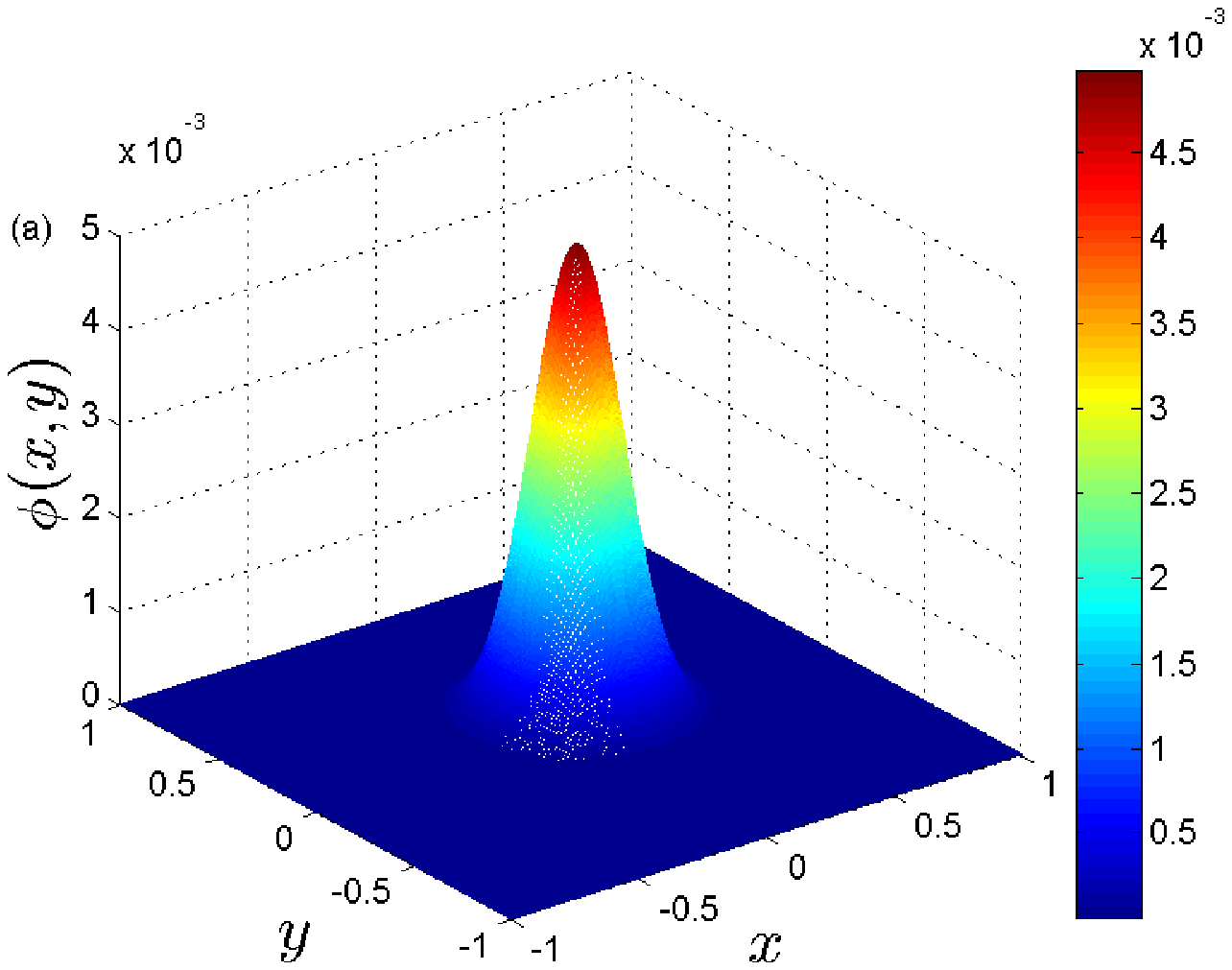}
\includegraphics[width=3in]{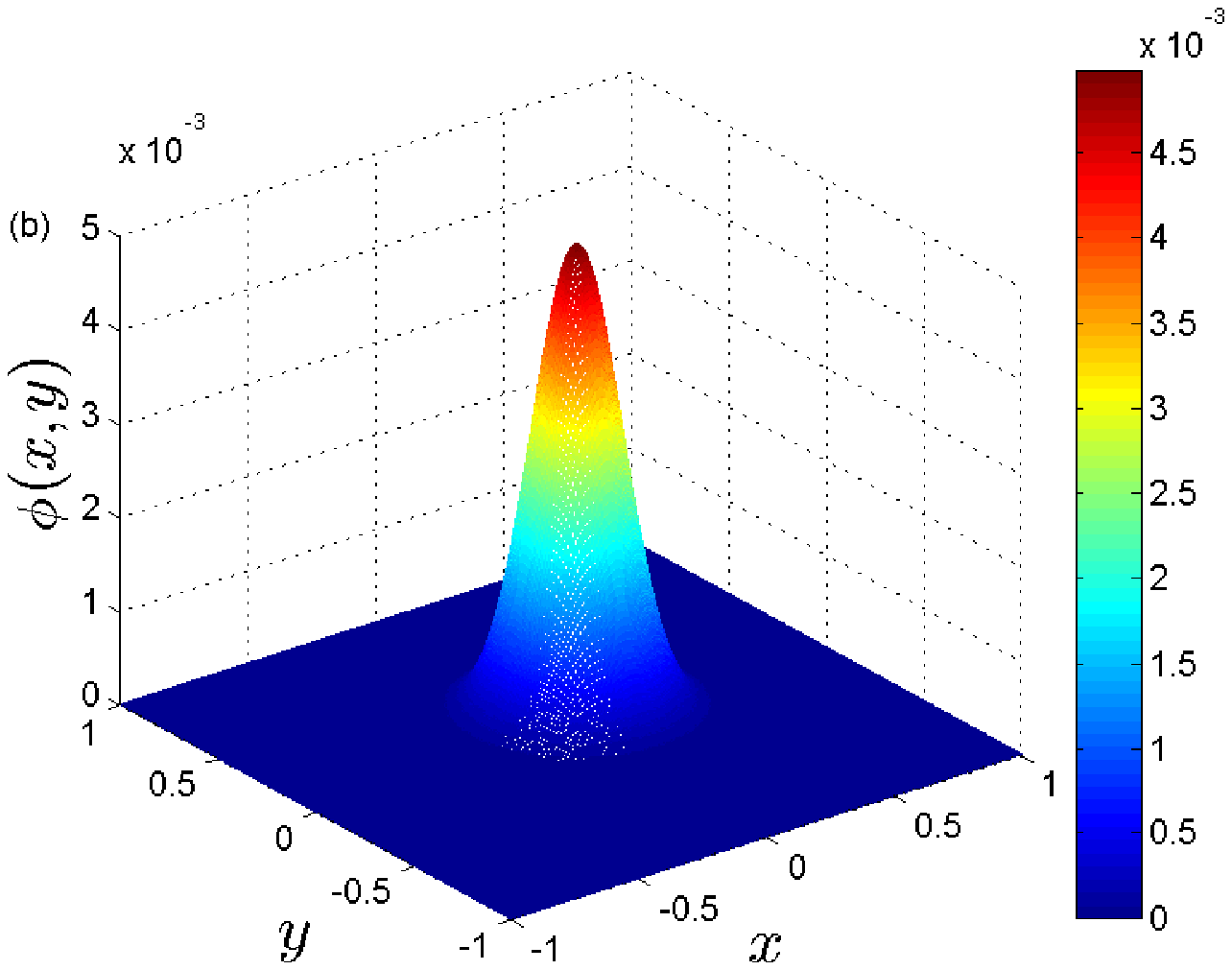}
\caption{\label{fig:8} Distributions of the scalar variable $\phi$ at the time $T=10$ [Isotropic diffusion problem: numerical solution (a), analytical solution (b)].}
\end{figure}

\begin{figure}
\includegraphics[width=3in]{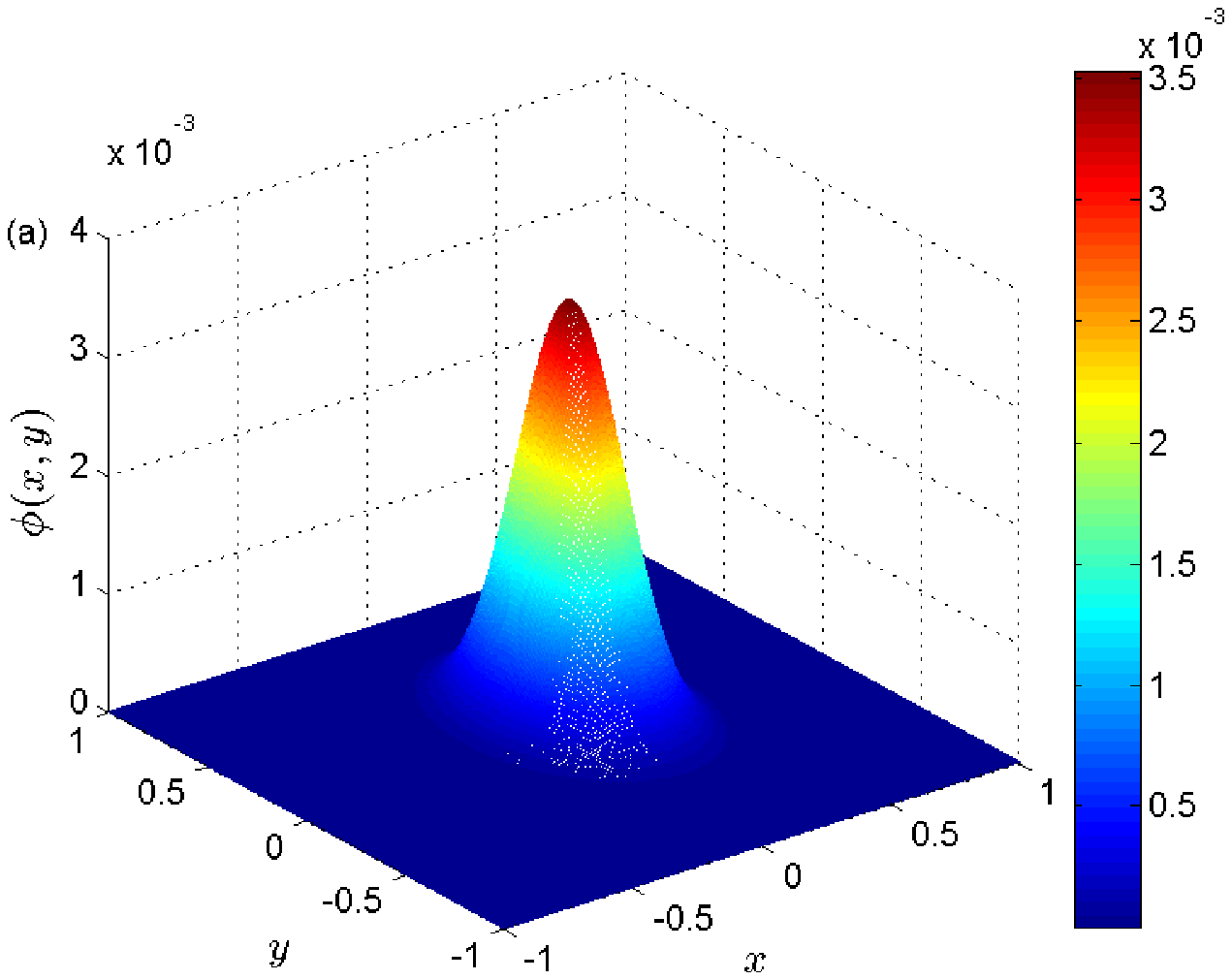}
\includegraphics[width=3in]{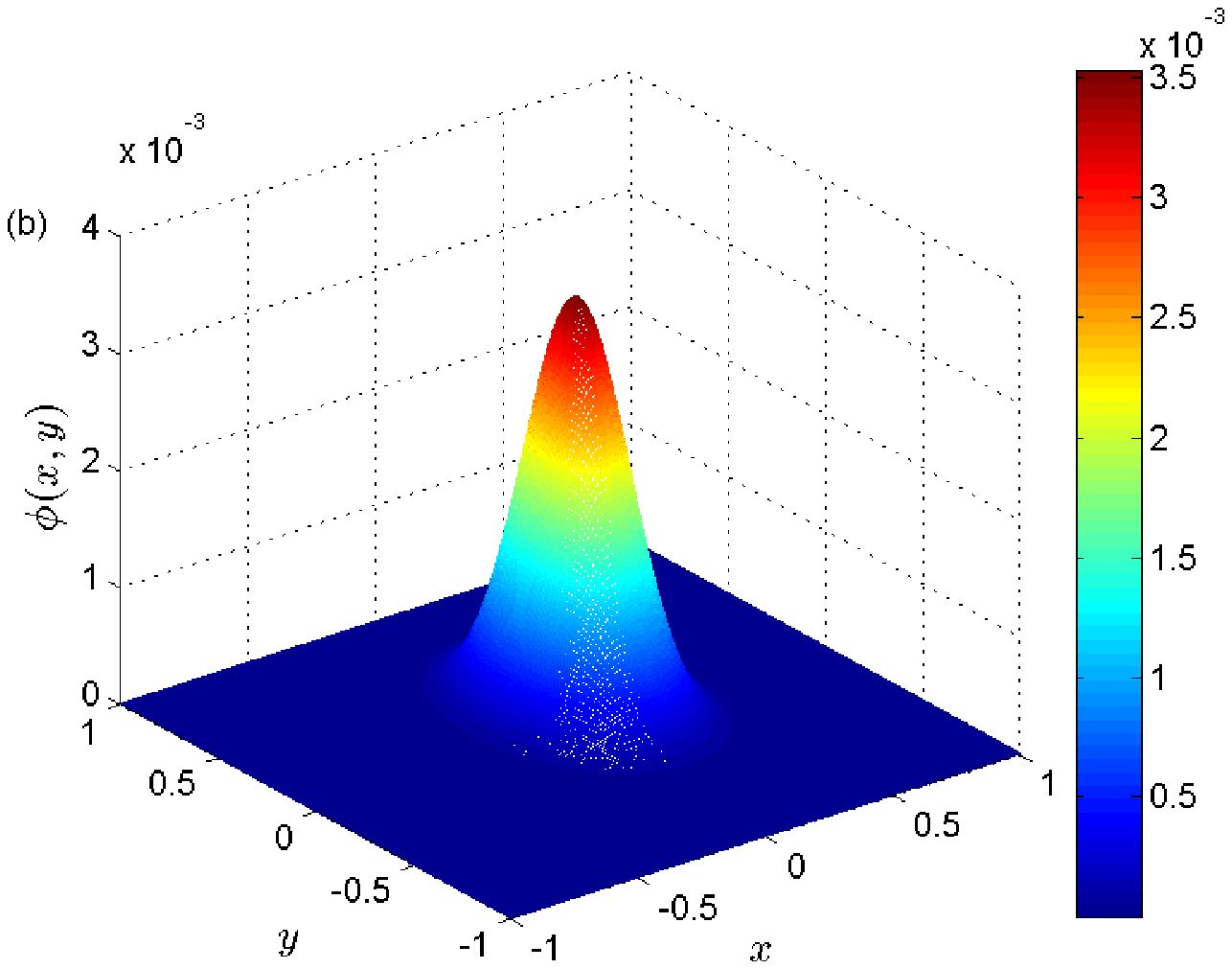}
\includegraphics[width=3in]{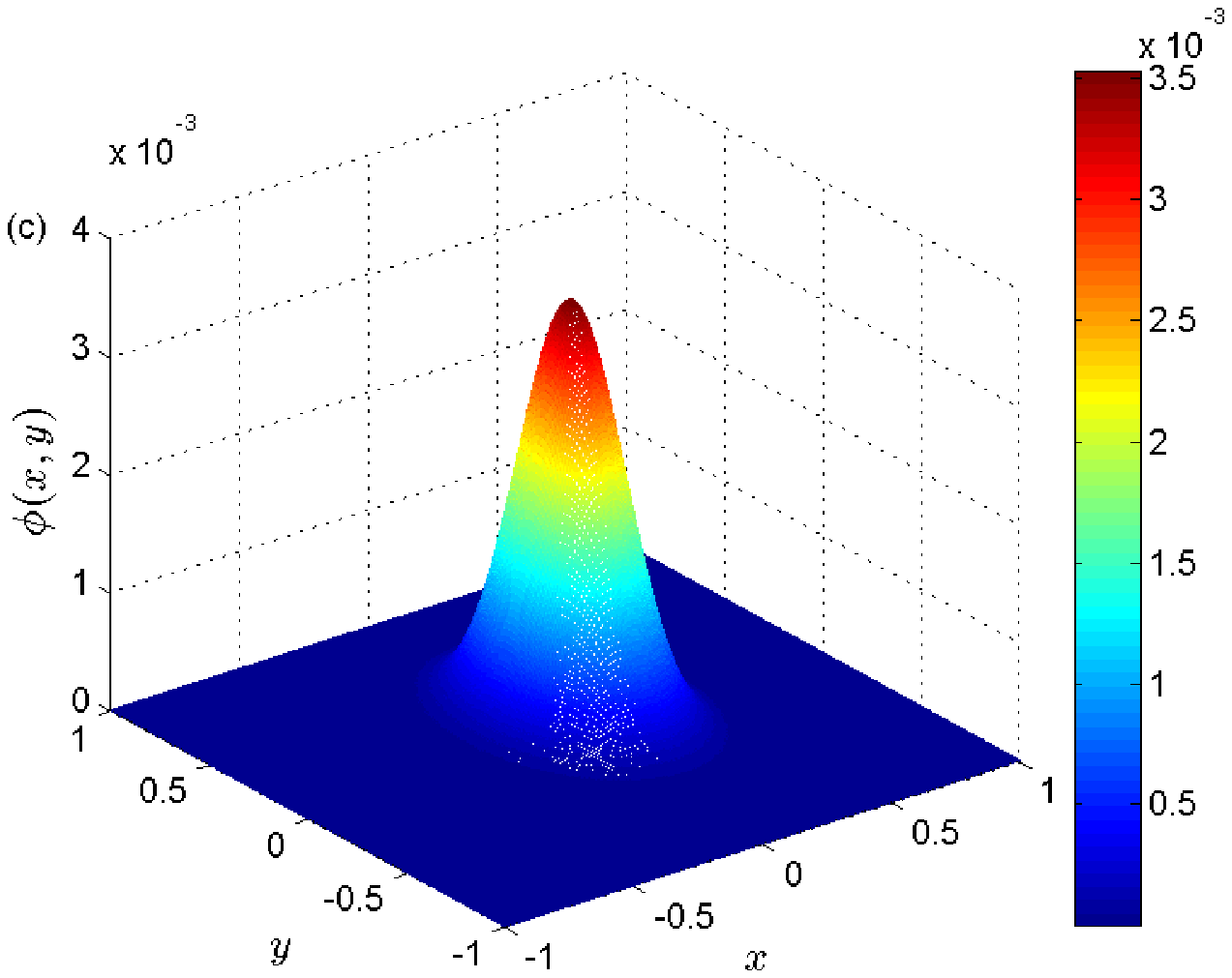}
\centering\caption{\label{fig:9} Distributions of the scalar variable $\phi$ at the time $T=10$ [Diagonally anisotropic diffusion problem: first approach (a), second approach (b), analytical solution (c)].}
\end{figure}

\begin{figure}
\includegraphics[width=3in]{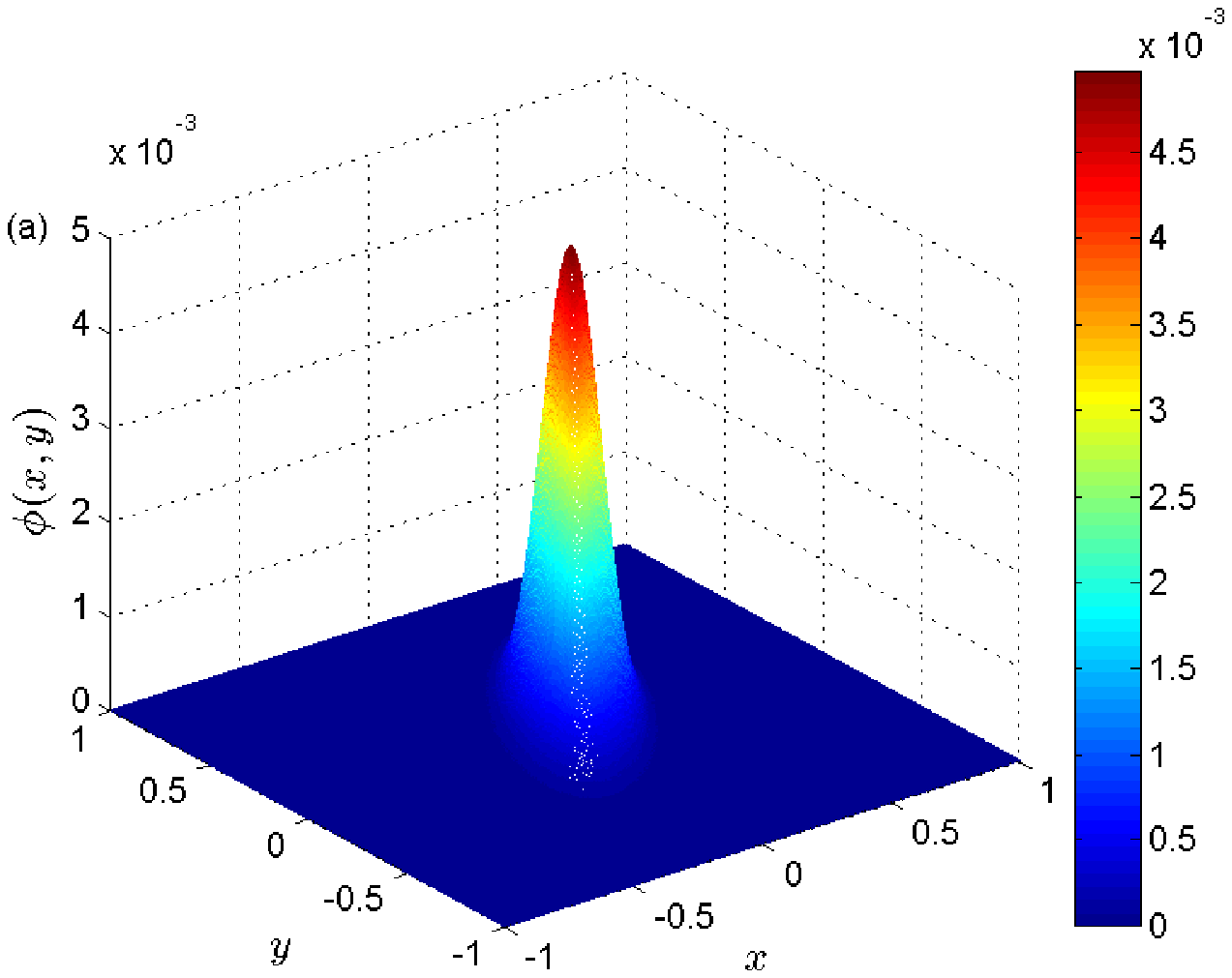}
\includegraphics[width=3in]{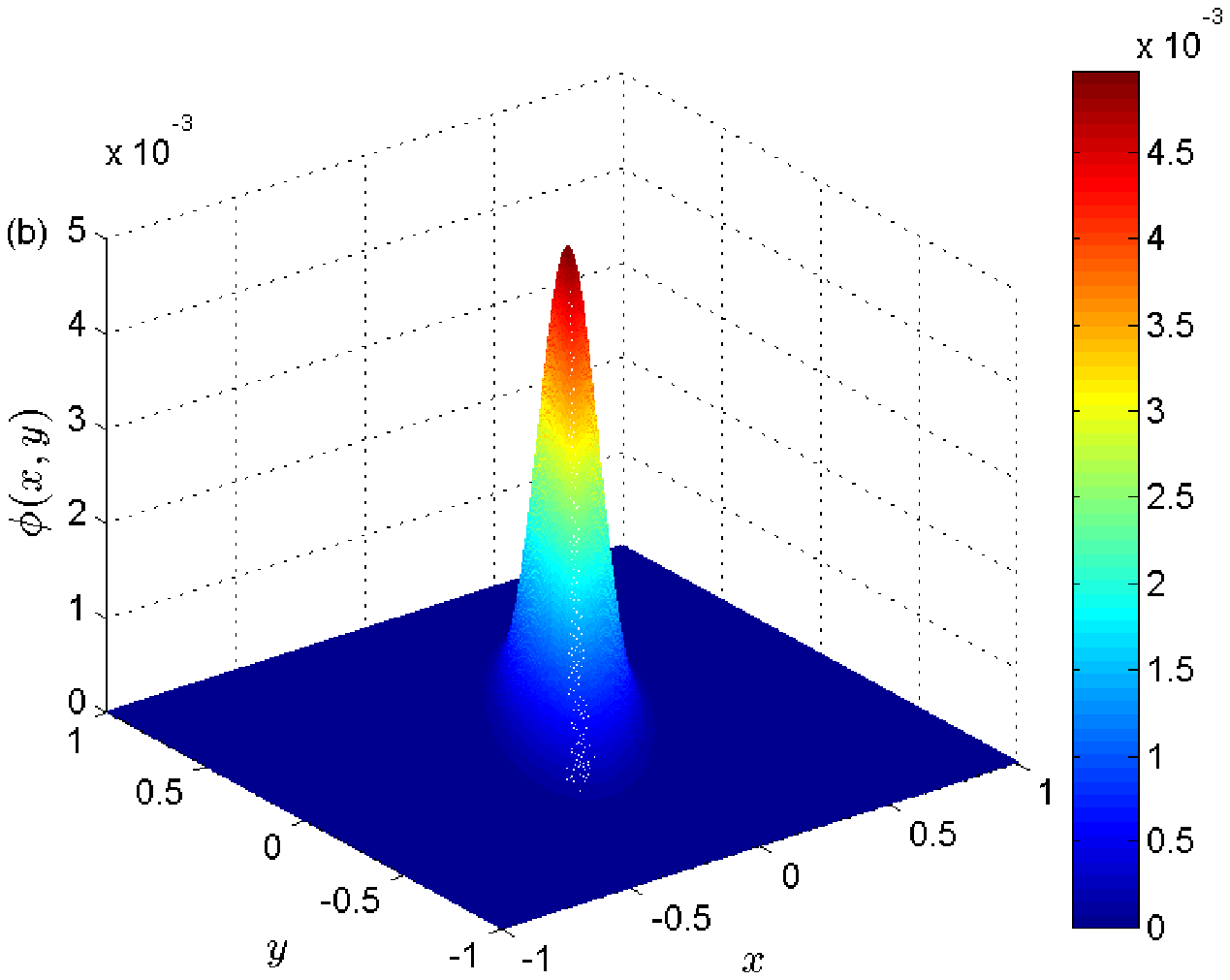}
\includegraphics[width=3in]{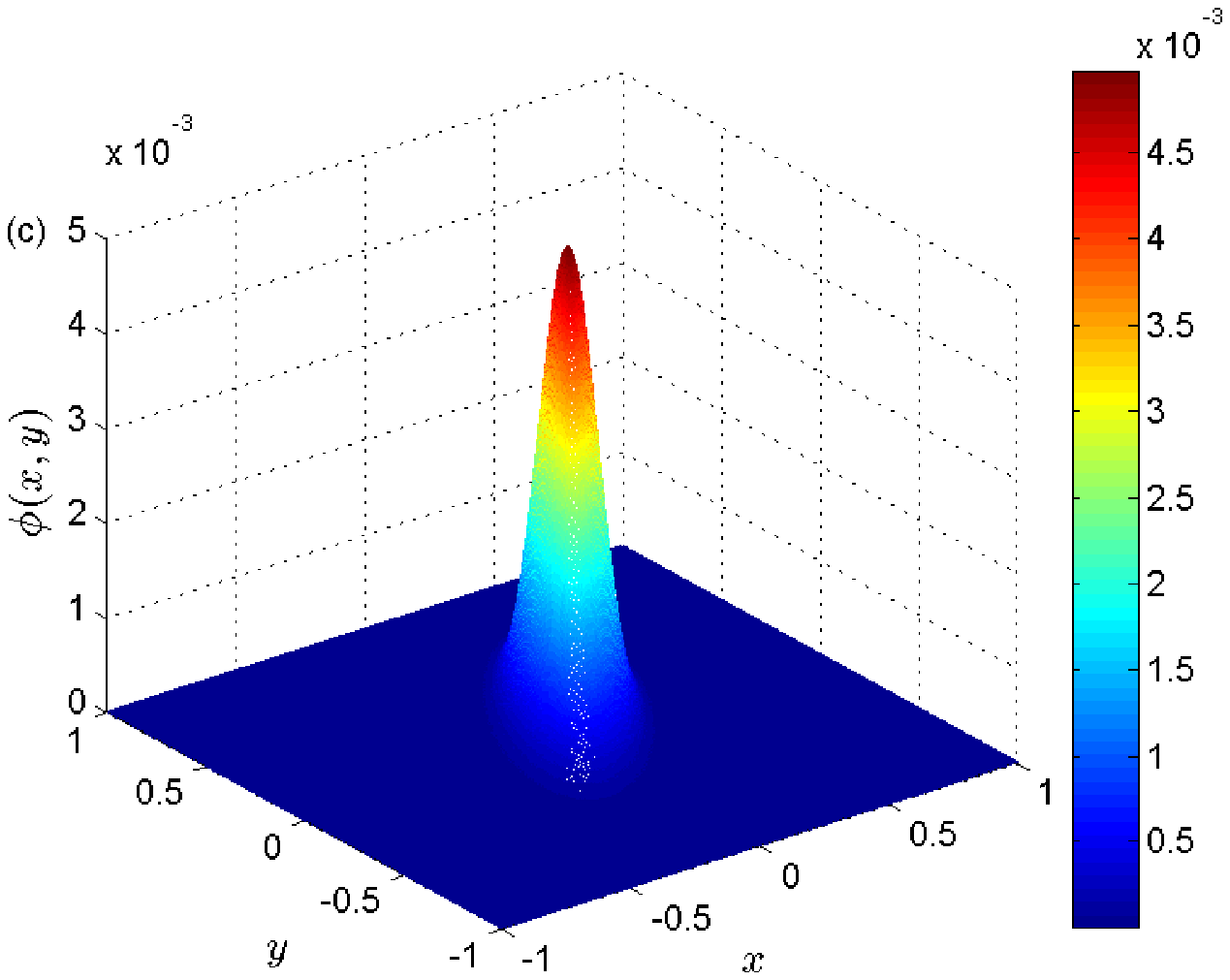}
\centering\caption{\label{fig:10} Distributions of the scalar variable $\phi$ at the time $T=10$ [Fully anisotropic diffusion problem: first approach (a), second approach (b), analytical solution (c)].}
\end{figure}

\begin{figure}
\includegraphics[width=3in]{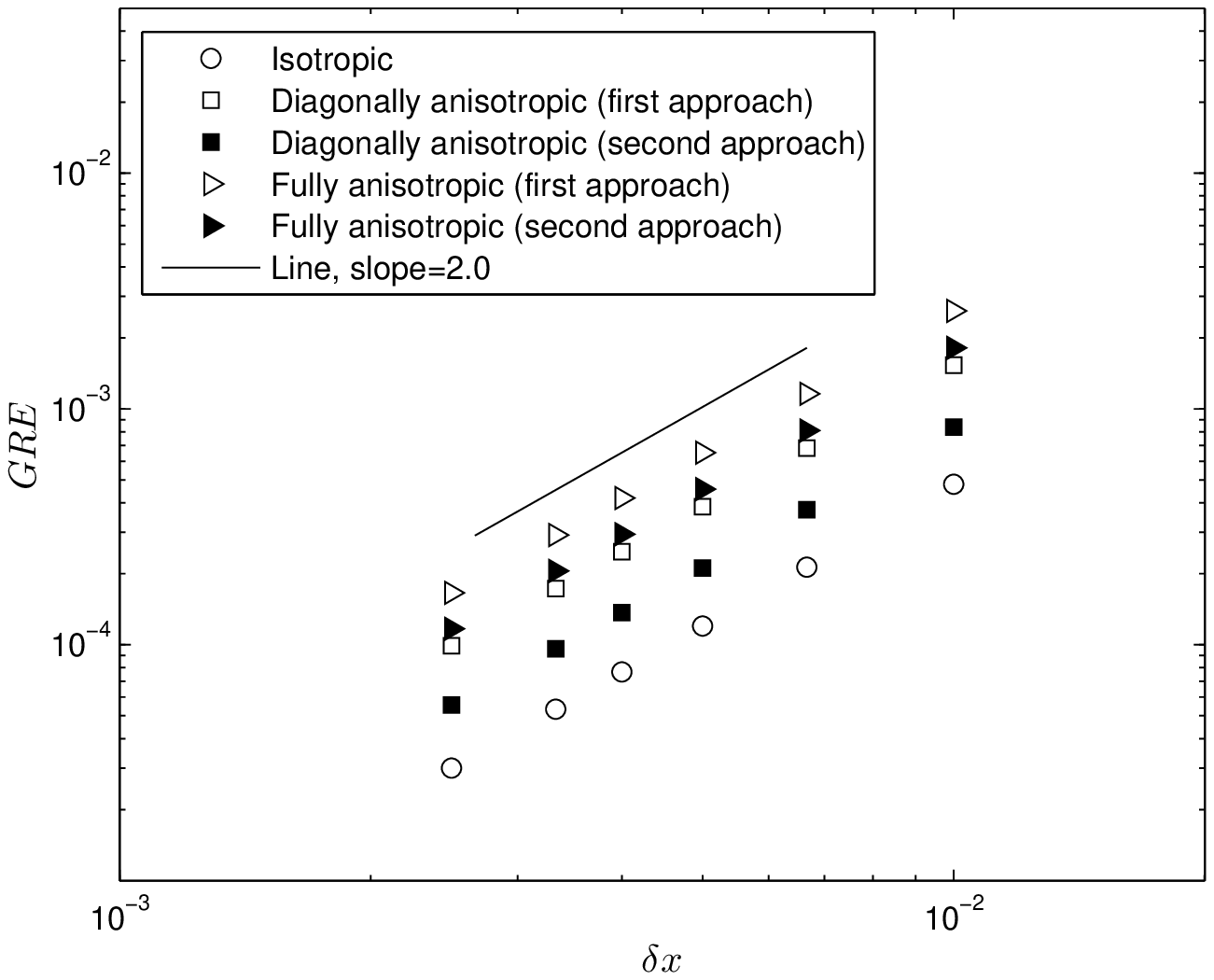}
\centering\caption{\label{fig:11} The global relative errors at
different lattice sizes ($\delta x=L/800$, $L/600$, $L/500$, $L/400$, $L/300$
and $L/200$, $L=2.0$ is the characteristic length), the slope of the inserted line is 2.0, which indicates that
the present MRT model has a second-order convergence rate in space.}
\end{figure}

\subsection{Anisotropic convection-diffusion
equation with constant velocity and variable diffusion tensor}

In this part, we will consider the following anisotropic CDE with a constant velocity $\mathbf{u}=(u_{x}, u_{y})^{\top}$ and a variable diffusion tensor $\mathbf{K}$,
\begin{equation}\label{eq64}
\partial_{t}\phi+\nabla\cdot(\phi\mathbf{u})=\nabla\cdot(\mathbf{K}\cdot\nabla\phi)+R.
\end{equation}
where $R$ is the source term. We note that the problem is more complicated since the diffusion tensor $\mathbf{K}$ is a function of space $\mathbf{x}$. For this reason, we cannot write Eq.~(\ref{eq64}) in an isotropic form, and thus the problem cannot be solved directly by the previous BGK model \cite{Shi2009}.

In this work, the diffusion tensor $\mathbf{K}$ is simply given by a diagonal matrix,
\begin{equation}\label{eq65}
\mathbf{K}=\kappa\left(
{\begin{array}{*{20}{c}}
   2-\sin(2\pi x)\sin(2\pi y) & 0  \\
   0 & 1 \\
\end{array}} \right),
\end{equation}
where $\kappa$ is a constant, and is fixed to be $1.0\times10^{-3}$. The source term $R$ is defined as
\begin{eqnarray}\label{eq66}
R & = & \exp[(1-12\pi^{2}\kappa)t]\{\sin(2\pi x)\sin(2\pi y)+4\kappa\pi^{2}\cos(4\pi x)\sin^{2}(2\pi y)\nonumber\\
& + & 2\pi[u_{x}\cos(2\pi x)\sin(2\pi y)+u_{y}\sin(2\pi x)\cos(2\pi y)]\}.
\end{eqnarray}
Under the periodic boundary conditions on the physical region $[0, 1]\times[0, 1]$ and the following initial condition,
\begin{equation}\label{eq67}
\phi(x, y, t=0)=\sin(2\pi x)\sin(2\pi y),
\end{equation}
one can derive the exact solution of the problem,
\begin{equation}\label{eq68}
\phi=\exp[(1-12\pi^{2}\kappa)t]\sin(2\pi x)\sin(2\pi y).
\end{equation}

Based on Eq.~(\ref{eq64}), one can determine the functions $\mathbf{B}$, $\mathbf{C}$ and $\mathbf{D}$, which are the same as those appeared in Eq.~(\ref{eq61}). We now performed some simulations with a fixed lattice size $401\times401$, and presented the results at time $T=3.0$ and different P\'{e}clet numbers (Pe=$Lu_{x}/\kappa$, $L=1.0$ is the characteristic
length, $u_{x}=u_{y}=0.1$ is the characteristic velocity) in Figs. 12 and 13 where Pe=100 and 1000. As seen from these figures, the numerical results are very close to the analytical solutions. To quantitatively measure the deviations between numerical results and corresponding analytical solutions, we also computed the $GRE$s of these two cases, and found that the values of them are $6.207\times10^{-4}$ and $9.831\times10^{-5}$, which are small enough and can be used to demonstrate that the present MRT model is accurate in the study of the anisotropic CDE with a variable diffusion tensor.

\begin{figure}
\includegraphics[width=3in]{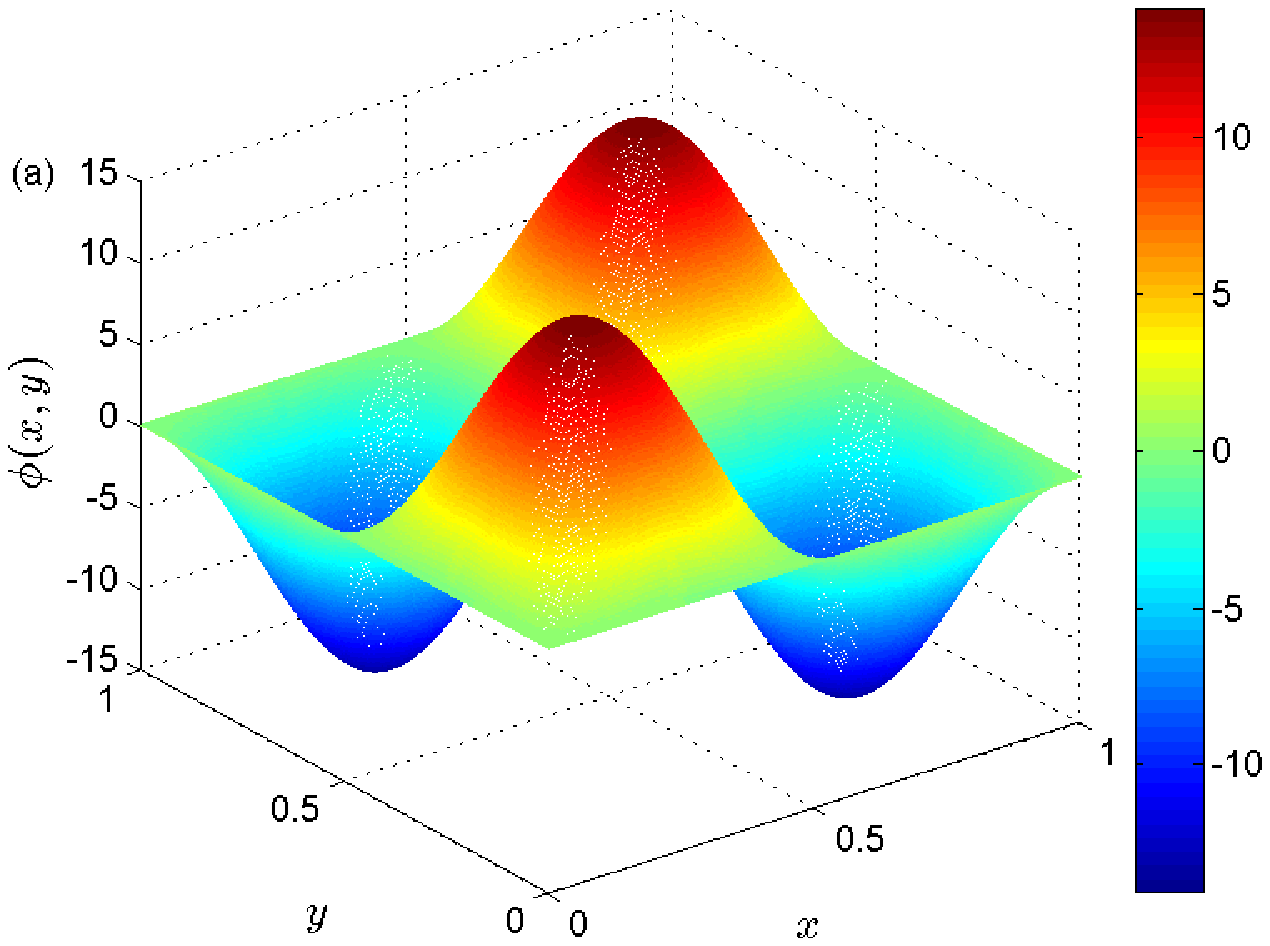}
\includegraphics[width=3in]{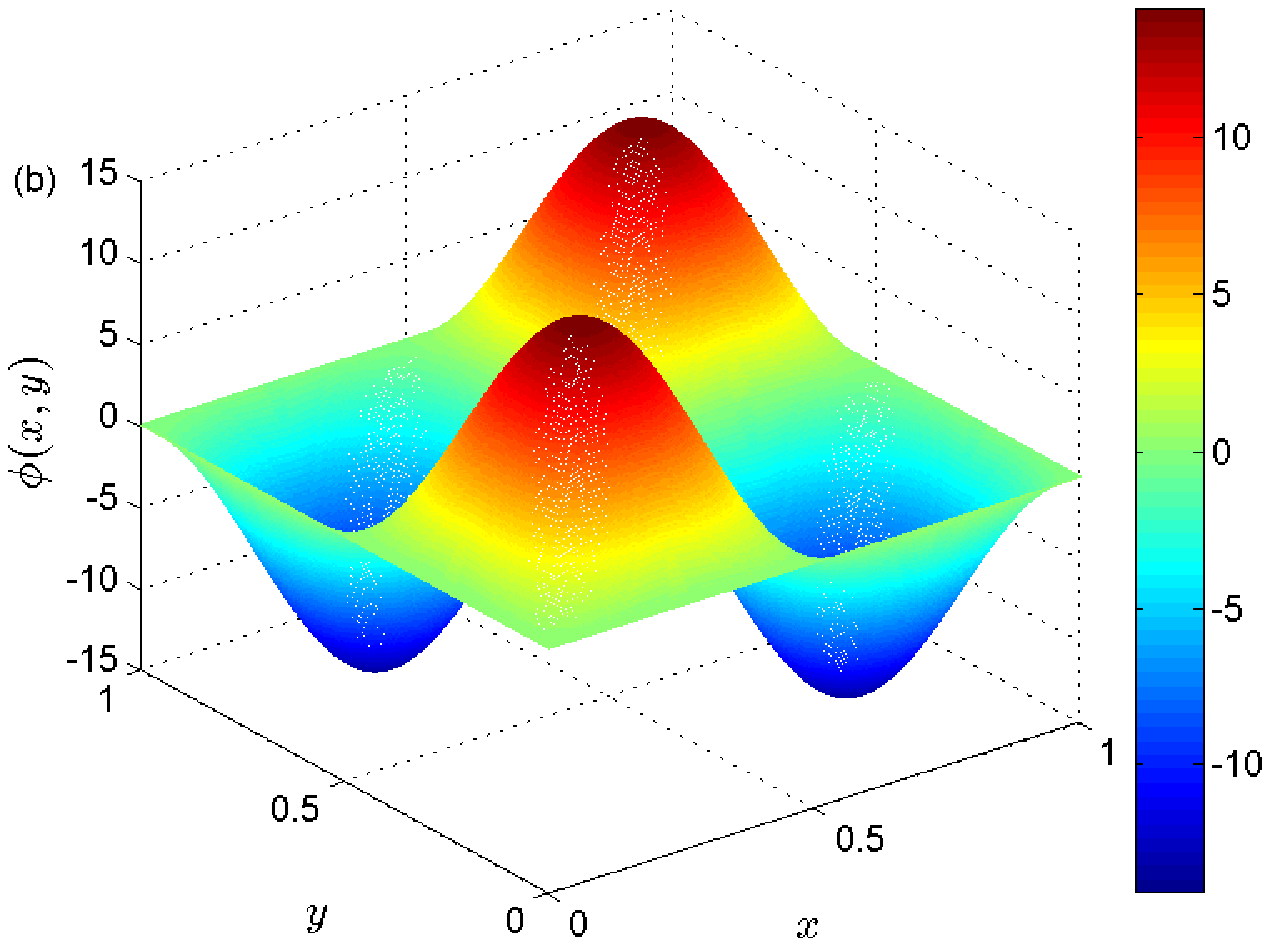}
\caption{\label{fig:12} Distributions of the scalar variable $\phi$ at the time $T=3$ and Pe=100 [(a): numerical solution, (b): analytical solution].}
\end{figure}

\begin{figure}
\includegraphics[width=3in]{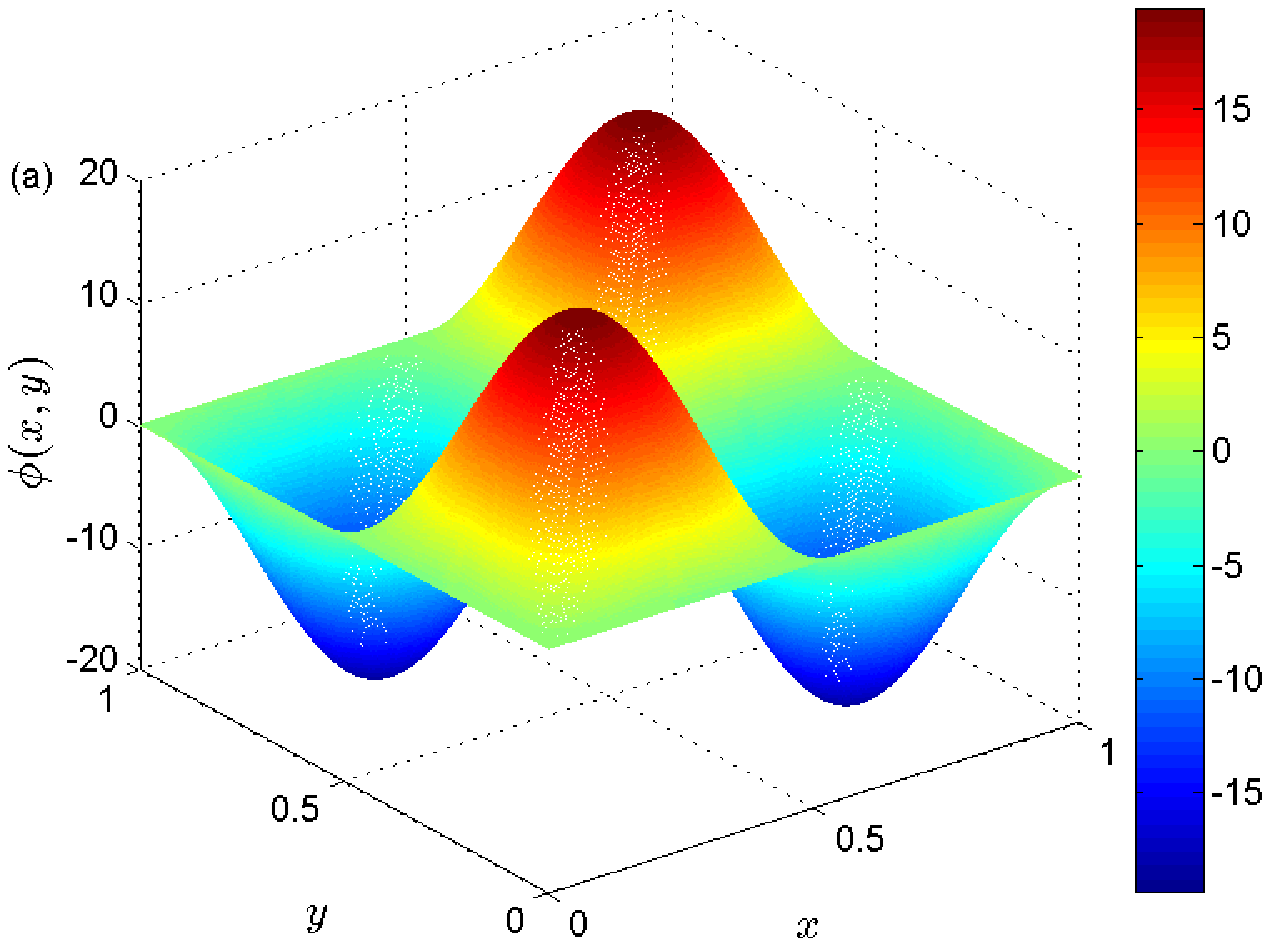}
\includegraphics[width=3in]{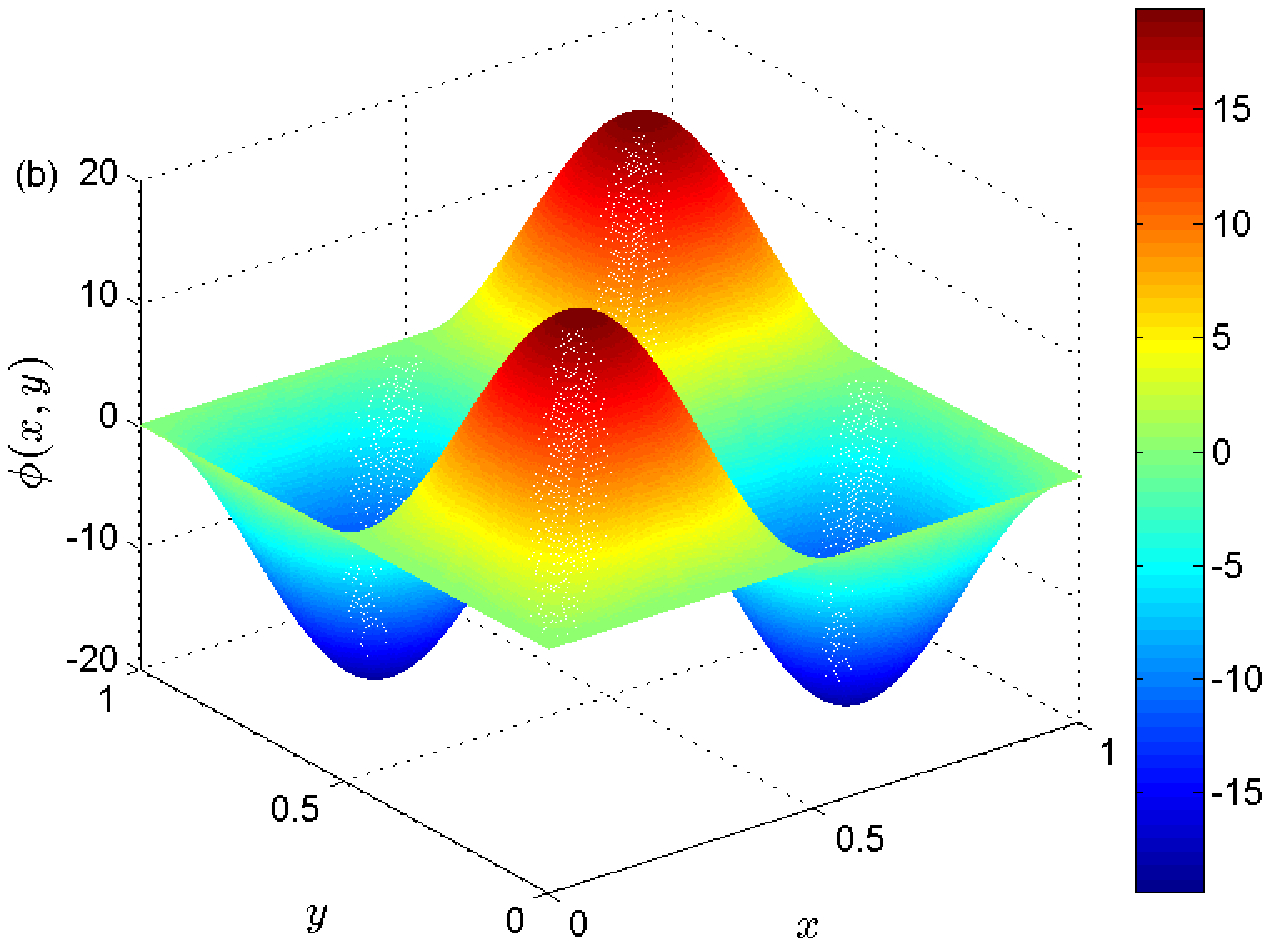}
\caption{\label{fig:13} Distributions of the scalar variable $\phi$ at the time $T=3$ and Pe=1000 [(a): numerical solution, (b): analytical solution].}
\end{figure}

To show the convergence rate of the present MRT for such complicated problem, we also carried out a number of simulations under different lattice resolutions, and presented the results in Fig. 14 where the lattice size is varied from $201\times201$ to $801\times801$. As shown in this figure, the present MRT model also has a second-order convergence rate for this special problem.

\begin{figure}
\includegraphics[width=3in]{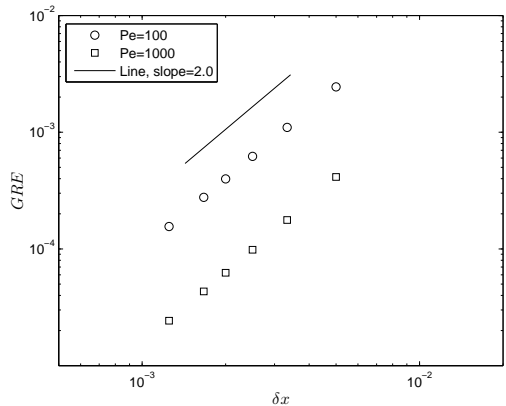}
\centering \caption{\label{fig:14} The global relative errors at
different lattice sizes ($\delta x=L/800$, $L/600$, $L/500$, $L/400$,
$L/300$ and $L/200$), the slope of the inserted line is 2.0, indicating that
the present MRT model has a second-order convergence rate in space.}
\end{figure}

\subsection{A comparison between the MRT model and BGK model}

 As reported in some available works \cite{Guo2013, Servan-Camas, Chai2014, Yoshida2010, Huang2014}, through tuning the relaxation parameters properly, the MRT model could be more accurate and more stable than the BGK model. To show the superiority of the MRT model over the BGK model, a comparison between two models is also conducted.

 We first performed a comparison of accuracy between the BGK model and MRT model through adopting a simple problem defined in a physical region $[0, L]\times[0, L]$, which can be described by the following CDE and boundary conditions,
\begin{subequations}\label{eq69}
\begin{equation}
\partial_{t}\phi+\nabla\cdot(\phi\mathbf{u})=\nabla\cdot(\kappa\nabla\phi)+R,
\end{equation}
\begin{equation}
\phi(t, x, 0)=\phi_{0}, \phi(t, x, L)=\phi_{L},
\end{equation}
\end{subequations}
where $\kappa$ is a constant diffusion coefficient, $\mathbf{u}=(u_{x}, u_{y})^{\top}$ is a constant velocity with $u_{y}=0$, $\phi_{0}$ and $\phi_{L}$ are two constants, $R=2\kappa\Delta\phi/L^{2}$ is the source term with $\Delta\phi=\phi_{L}-\phi_{0}$. Under an assumption that the problem is steady and unidirectional, i.e., $\phi$ is only a function of $y$, we can derive analytical solution of the problem,
\begin{equation}\label{eq70}
\phi(y)=\phi_{0}+\Delta\phi\frac{y}{L}(2-\frac{y}{L}).
\end{equation}
The reason for choosing this problem is that, following a similar procedure reported in Ref. \cite{He1997}, one can readily derive the analytical solutions of the BGK model and MRT model (\textbf{Scheme B} with $s_{1}=s_{2}=s_{4}=s_{6}=s_{7}=s_{8}$) with adopting the anti-bounce back boundary condition \cite{Ginzburg2005b, Zhang2012, Chai2014},
\begin{equation}\label{eq71}
\phi_{j}=\phi_{0}-\Delta\phi \bar{y}_{j}(2.0-\bar{y}_{j})+\phi_{s},
\end{equation}
where $\bar{y}_{j}=(j-1/2)/N$ with $N$
 representing the grid number used in $y$ direction, $\phi_{s}$ is \emph{numerical} slip caused by the model adopted, and can be given by
\begin{subequations}\label{eq72}
\begin{equation}
\phi_{s, BGK}=\frac{\Delta\phi}{12N^{2}}[4(\frac{2}{s_{BGK}}-1)^{2}-3],
\end{equation}
\begin{equation}
\phi_{s, MRT}=\frac{\Delta\phi}{12N^{2}s_{1}s_{3}}[s_{1}s_{3}-8(s_{1}+s_{3})+16].
\end{equation}
\end{subequations}
Based on Eq.~(\ref{eq72}a), one can find that although the BGK model has a second-order convergence rate for this simple problem, the \emph{numerical} slip of the scalar variable $\phi$ cannot be eliminated unless $s_{BGK}=4(2-\sqrt{3})$. While for the MRT model, we can make the solution of MRT model [Eq.~(\ref{eq72}b)] consistent with that of the physical problem [Eq.~(\ref{eq70})] through setting $\phi_{s}$ to be zero, which means that the relaxation parameters $s_{1}$ and $s_{3}$ should satisfy the following relation,
\begin{equation}\label{eq73}
s_{1}=\frac{8(s_{3}-2)}{8-s_{3}}.
\end{equation}
From above theoretical analysis, it is clear that the MRT model can be more accurate than the BGK model through tuning the free relaxation parameter $s_{1}$. In addition, similar to the procedure used in above examples, the functions $\mathbf{B}$, $\mathbf{C}$ and $\mathbf{D}$, the diffusion tensor $\mathbf{K}$, and the source term $R_{i}$ used in our model can also be determined,
\begin{equation}\label{eq74}
\mathbf{B}=\phi\mathbf{u},\ \  \mathbf{C}=\phi\mathbf{u}\mathbf{u}, \ \ \mathbf{D}=\phi\mathbf{I},\ \ \mathbf{K}=\kappa\mathbf{I},\ \  R_{i}=\omega_{i}(1+\frac{\mathbf{c}_{i}\cdot\mathbf{u}}{c_{s}^{2}})R.
\end{equation}
To validate above analysis and confirm our statement, we also performed some simulations with different lattice resolutions and relaxation parameters ($s_{3}$), and presented the results in Fig. 15 where $L=1.0$, $u_{x}=0.1$, $\phi_{0}=0$, $\phi_{L}=1.0$, and the diffusion coefficient $\kappa$ is set to be 0.1.
As seen from Fig. 15, the numerical results obtained by the MRT model are in good agreement with the analytical solution [Eq.~\ref{eq70}] even with a coarse grid (e.g., $N=4$), while the results given by the BGK model deviate from the analytical solution unless the relaxation parameter $s_{BGK}$ is fixed to be $s_{3}=4(2-\sqrt{3})$, which is consistent with above theoretical analysis.

\begin{figure}
\includegraphics[width=3in]{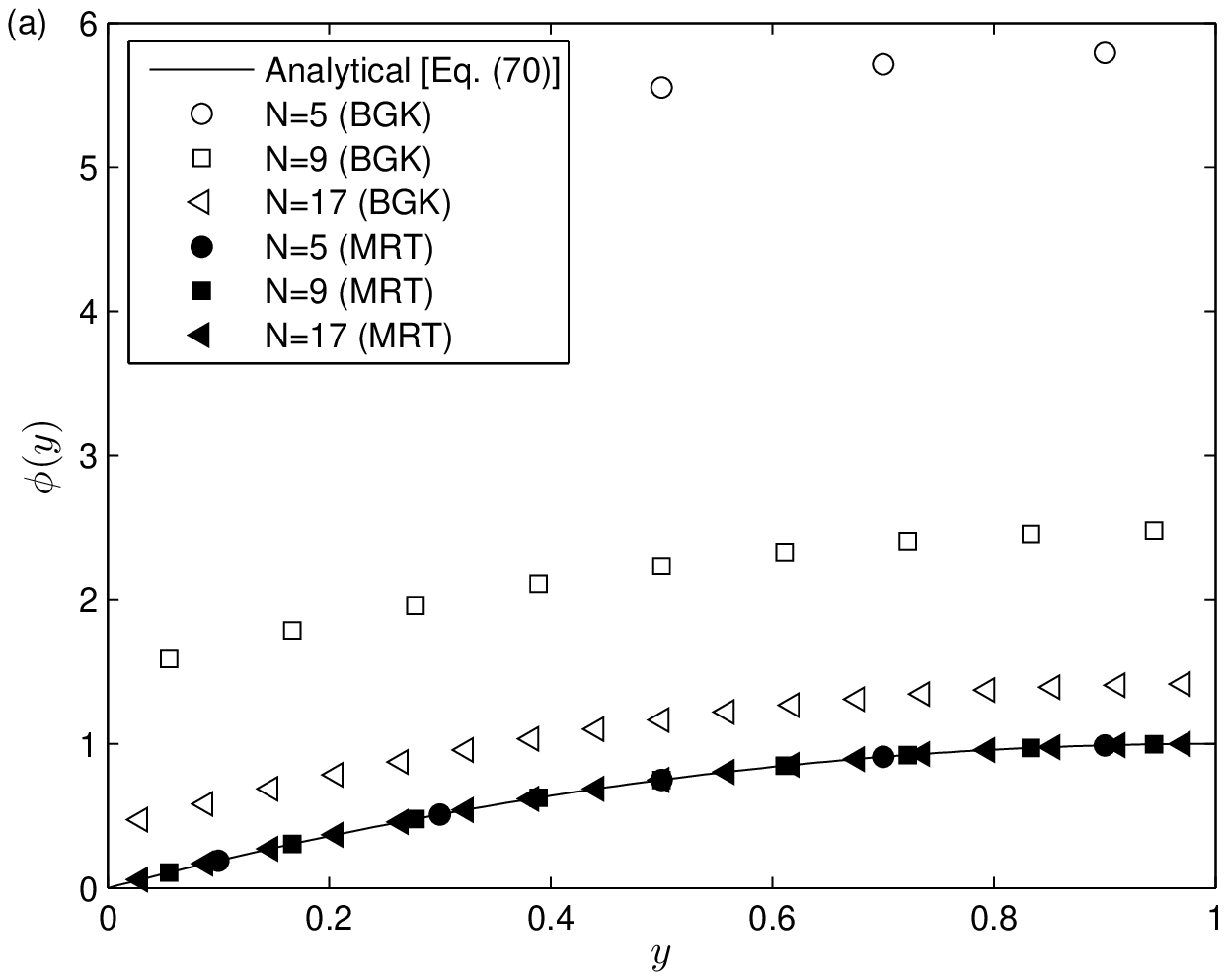}
\includegraphics[width=3in]{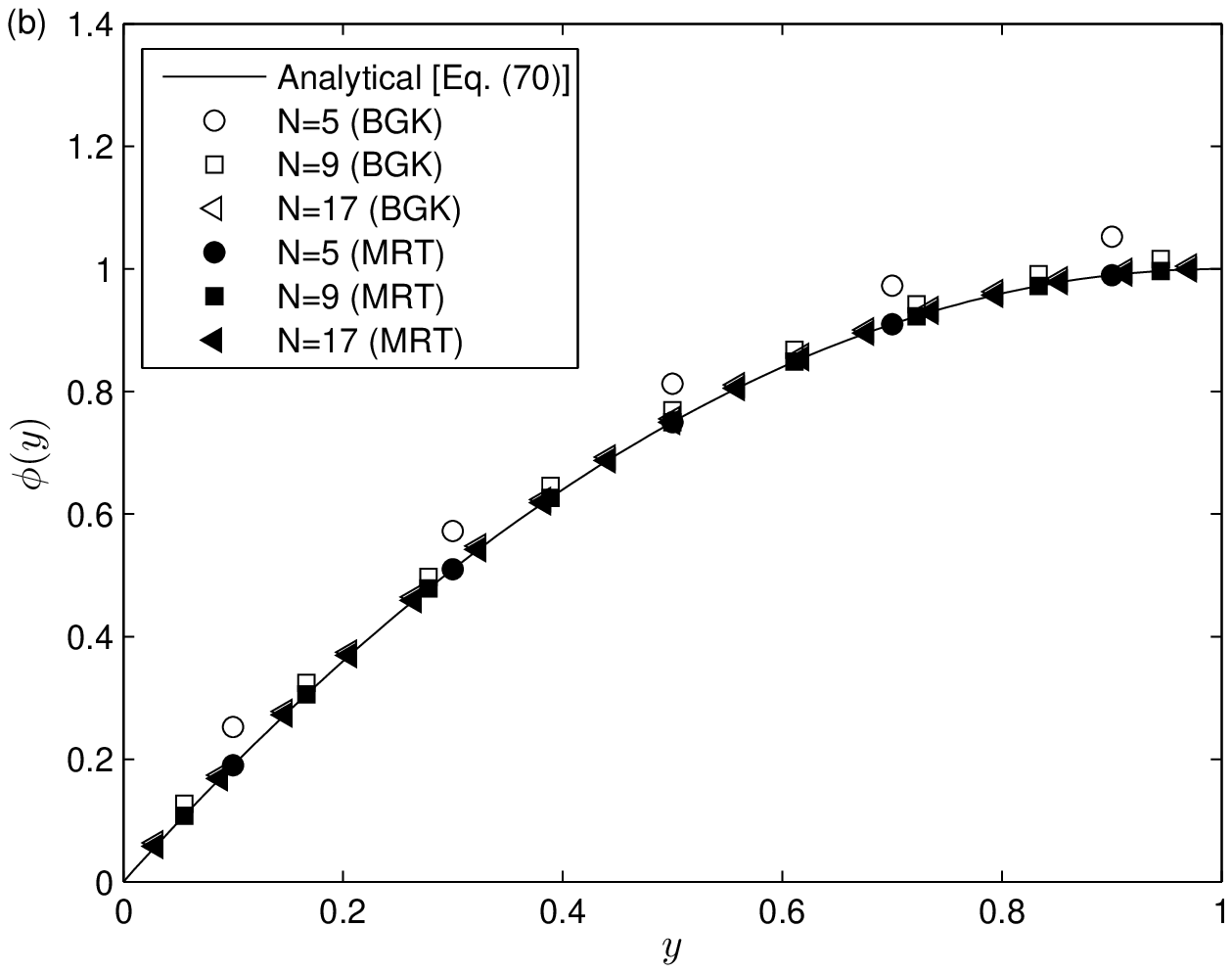}
\includegraphics[width=3in]{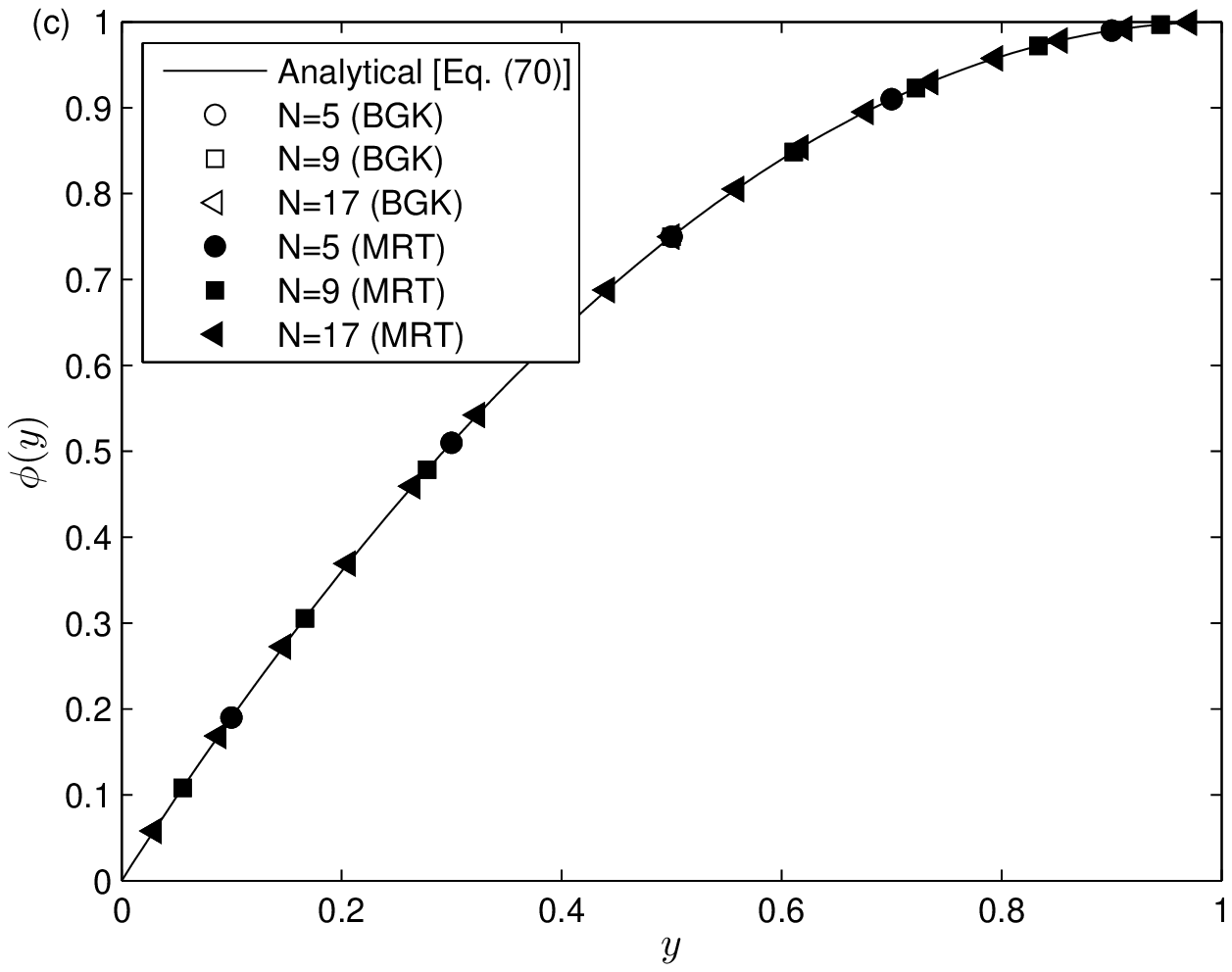}
\includegraphics[width=3in]{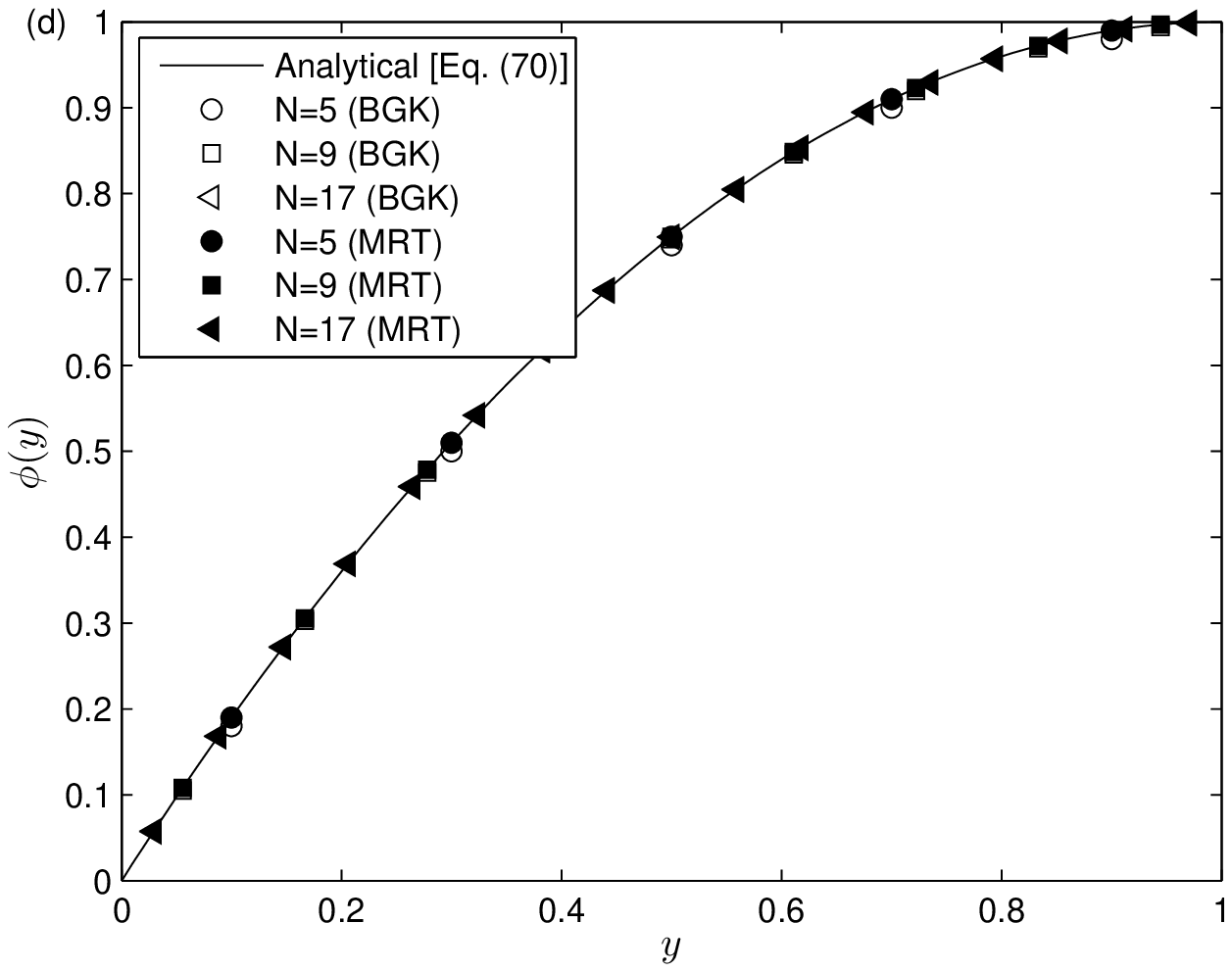}
\caption{\label{fig:15} Comparisons between the BGK model and MRT model under different lattice sizes and relaxation parameters [(a): $s_{3}=0.1$, (b): $s_{3}=0.6$, (c): $s_{3}=4(2-\sqrt{3})$, (d): $s_{3}=1.9$].}
\end{figure}

Then a comparison of stability between the BGK model and MRT model is also carried out through using the Gaussian hill problem, which has been investigated previously. Here we only take the fully anisotropic diffusion problem as an example, while to simply perform a comparison between the BGK and MRT models, the second approach presented in section \emph{3.4} [the anisotropic CDE is written in an isotropic form, see Eq.~(\ref{eq62})] is adopted. In the following simulations,
the diffusion tensor $\mathbf{K}$ is taken as
\begin{equation}\label{eq75}
\mathbf{K}=\kappa\left(
{\begin{array}{*{20}{c}}
   1 & 1  \\
   1 & 2 \\
\end{array}} \right),
\end{equation}
from which one can further determine the function $\mathbf{D}=\mathbf{K}\phi/\kappa$ in the second approach [see Eq.~(\ref{eq61})]. $\kappa$ is a constant, and is varied to test the stability of the BGK model and MRT model. The lattice speed $c$ is equal to 1.0, the velocity $\mathbf{u}$ is fixed to be $\mathbf{u}=(0.1, 0.1)^{\top}$. For the relaxation parameters appeared in the MRT model, $s_{3}$ and $s_{5}$ can be determined from $\kappa$, and the others are fixed through the following equation,
\begin{equation} \label{eq76}
 s_{0}=0,\ s_{1}=s_{2}=s_{4}=s_{6}=s_{7}=s_{8}=s,
\end{equation}
where $s$ can be varied in a proper range to ensure that the MRT model is stable, but for simplicity, only a special case $s=0.6$ is considered. The other parameters used in simulations are the same as those appeared in section \emph{3.4} except for time $T=5.0$.

We first conducted simulations with $\kappa=10^{-3}$, and presented the results in Fig.16. As shown in this figure, both numerical results obtained by the BGK model and MRT model are close to the analytical solution. However, when $\kappa$ is decreased to $10^{-4}$, the BGK model is unstable, while the MRT model can give a stable solution (see Fig. 17), and the $GRE$ is about $4.438\times10^{-2}$.
\begin{figure}
\includegraphics[width=3in]{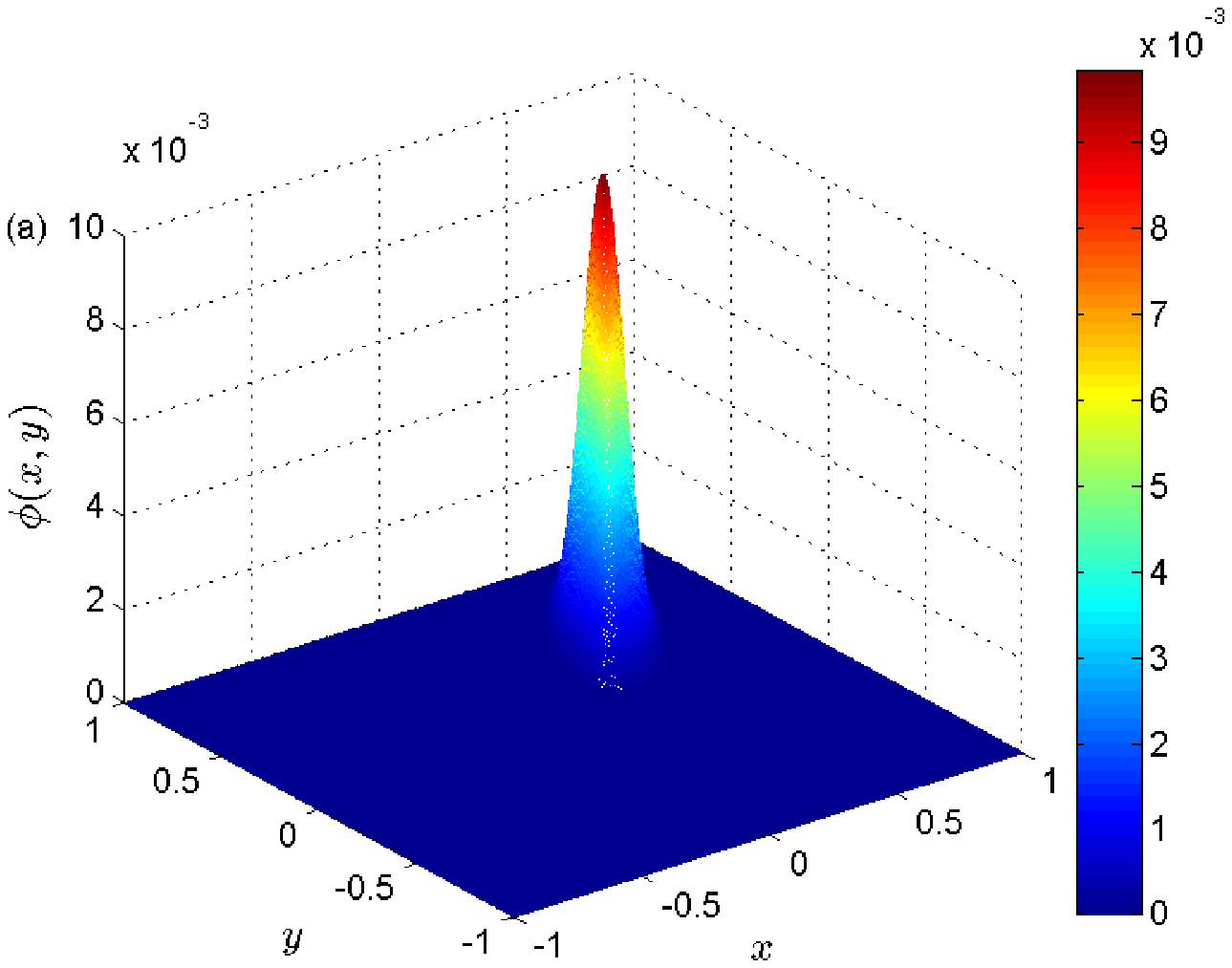}
\includegraphics[width=3in]{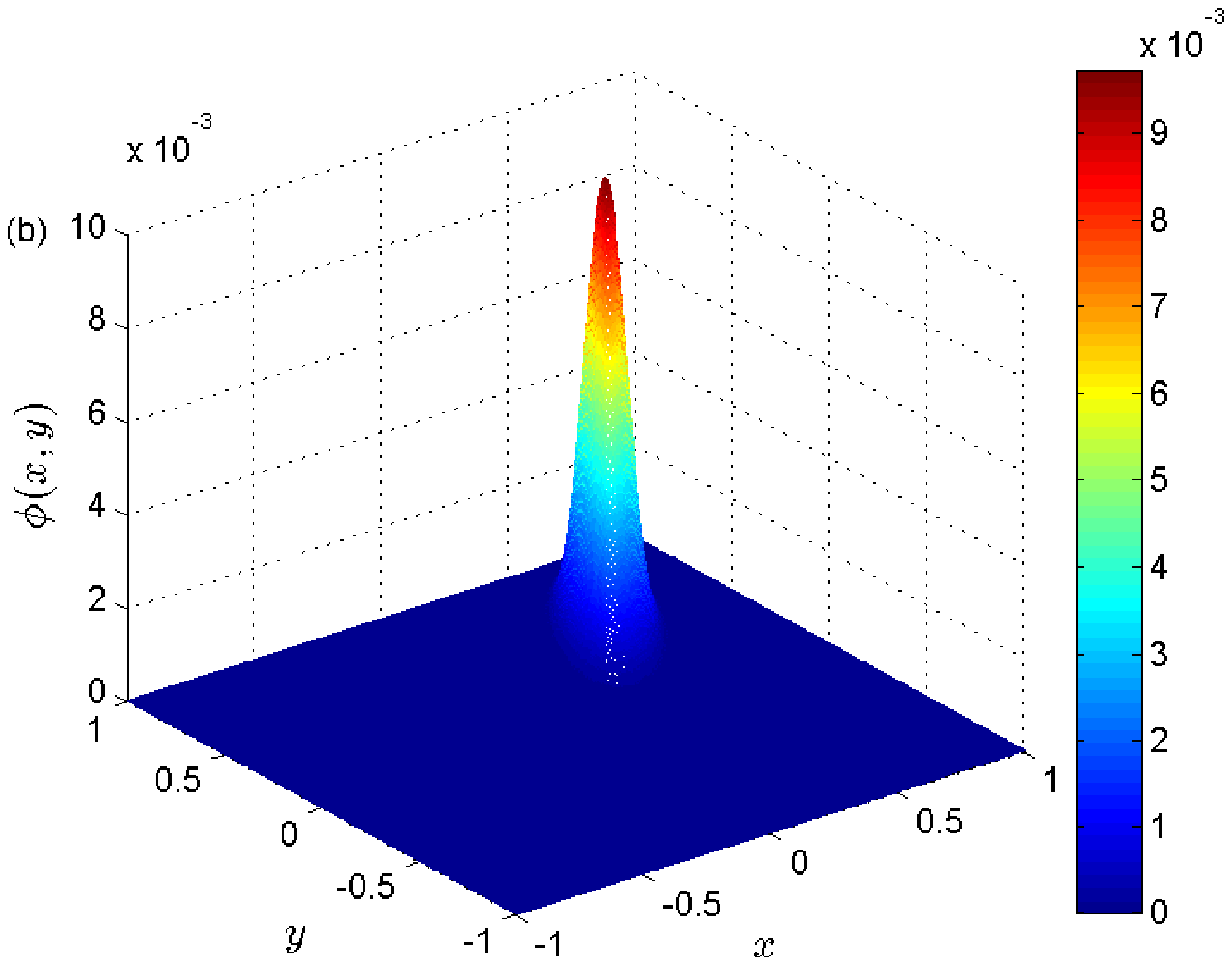}
\includegraphics[width=3in]{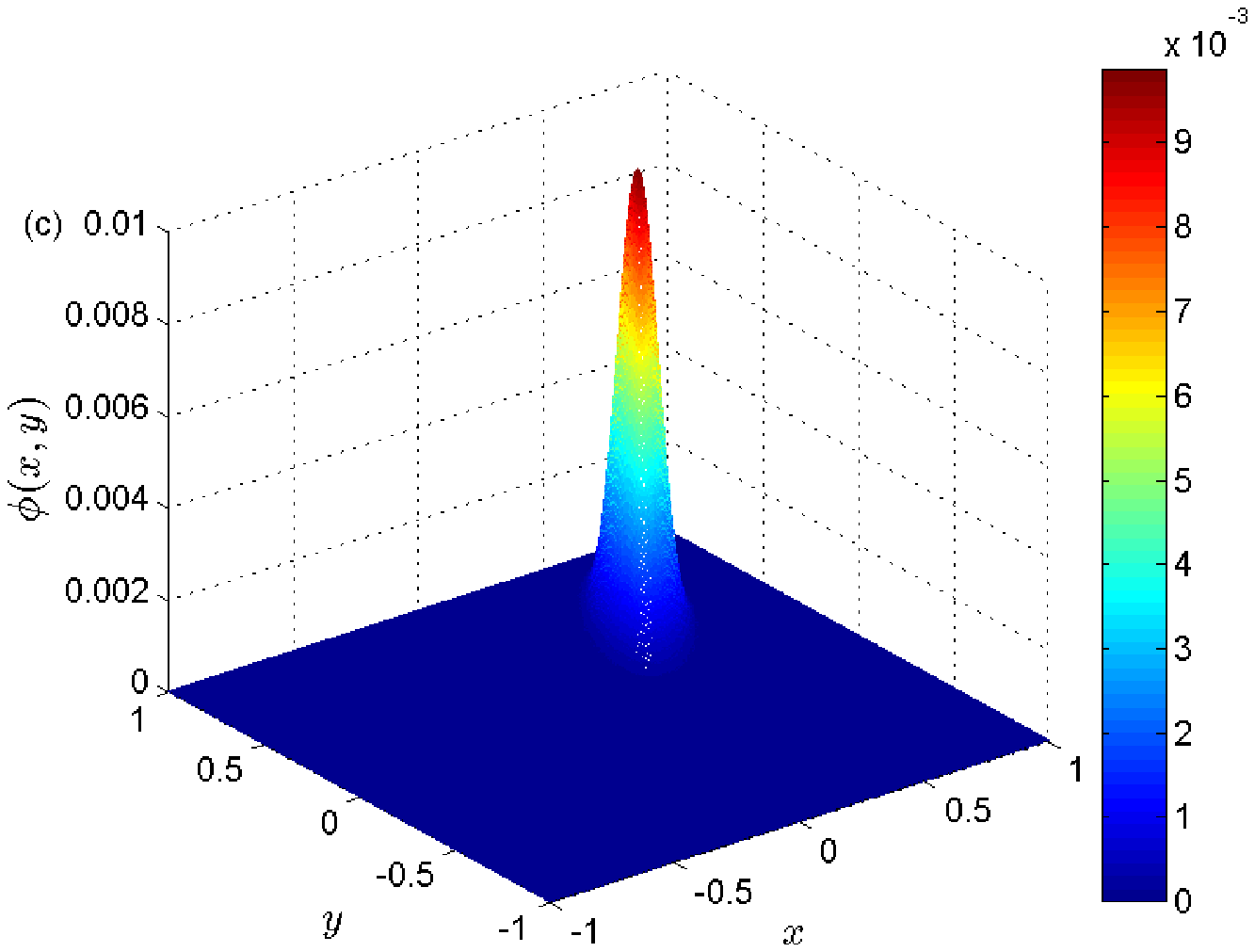}
\centering\caption{\label{fig:16} A comparison between the BGK model and MRT model for fully anisotropic diffusion problem ($\kappa=10^{-3}$) [(a): BGK model, (b): MRT model, (c): Analytical solution].}
\end{figure}
\begin{figure}
\includegraphics[width=3in]{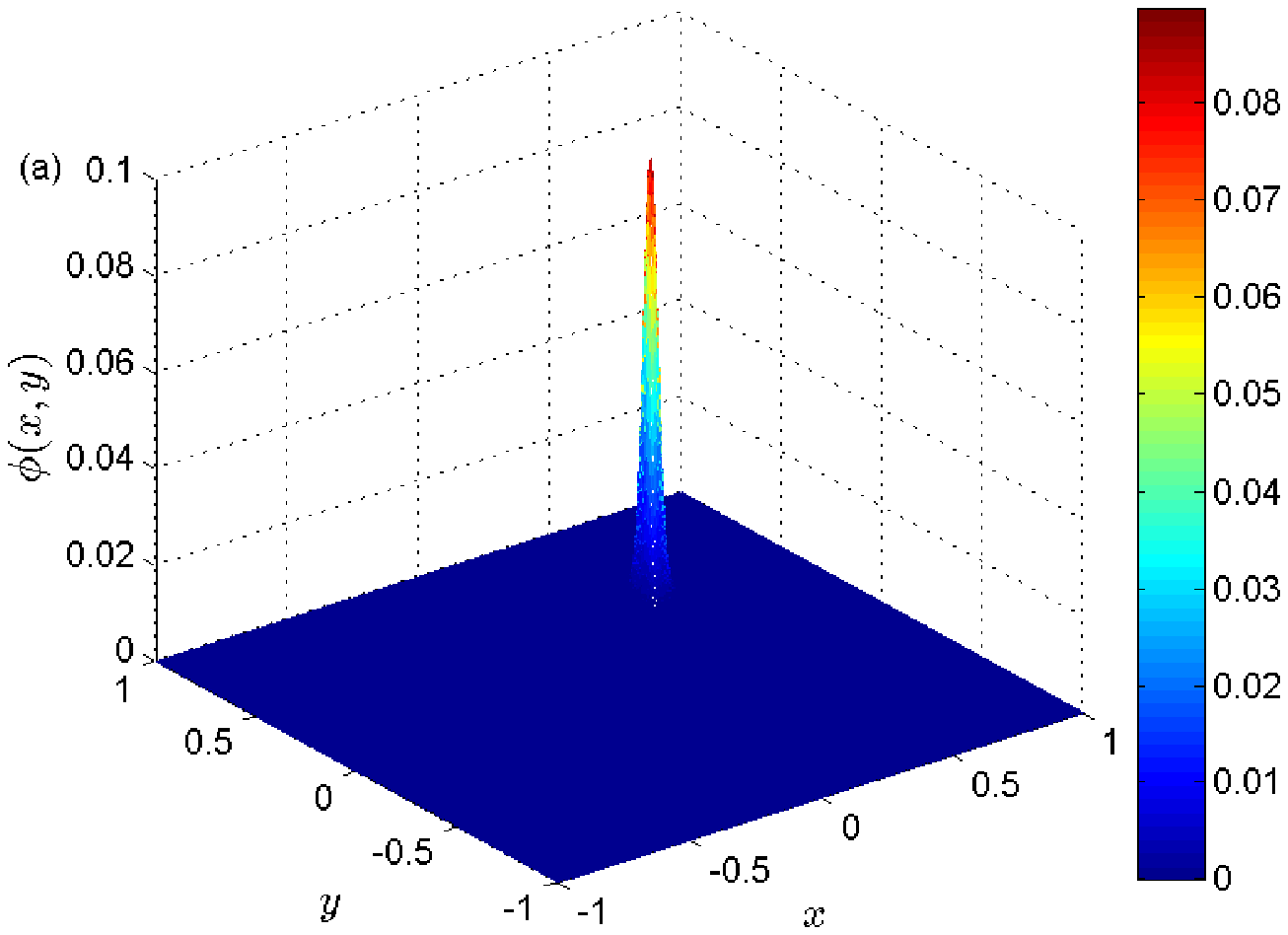}
\includegraphics[width=3in]{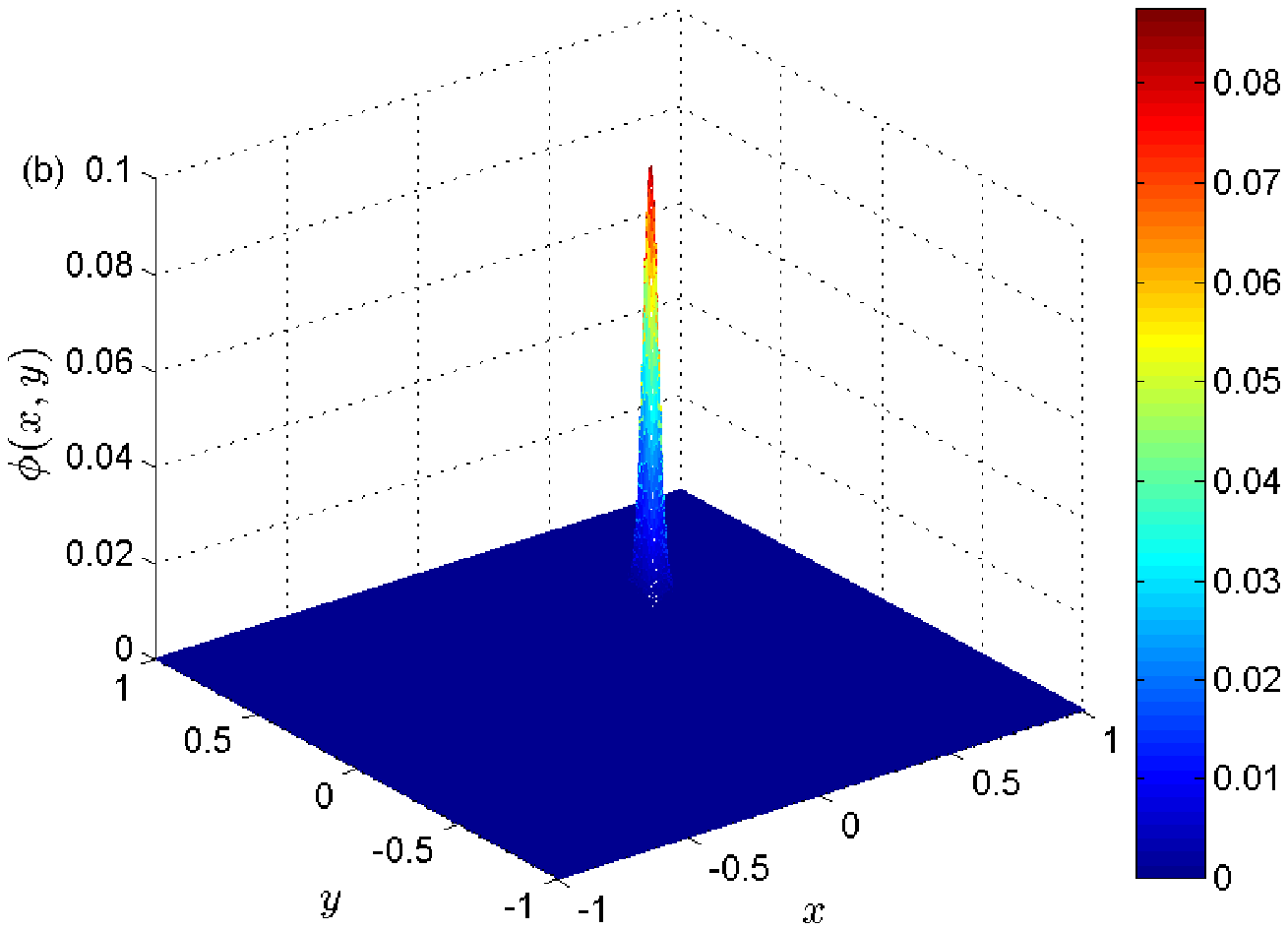}
\caption{\label{fig:17} A comparison between the BGK model and MRT model for fully anisotropic diffusion problem ($\kappa=10^{-4}$) [(a): MRT model, (b): Analytical solution].}
\end{figure}
From above discussion, it is clear that, through tuning the relaxation parameters properly, the MRT could be more accurate and more stable than the BGK model.

\section{Conclusions}

In this work, a general multiple-relaxation-time lattice
Boltzmann model for nonlinear anisotropic convection-diffusion equations is proposed, and is then tested by some
classic NACDEs, including the simple linear CDE,
nonlinear Burgers-Fisher equation, nonlinear Buckley-Leverett equation and
some anisotropic CDEs. The numerical
results show that the present MRT model is efficient and accurate in solving the NACDEs, and also has a second-order convergence rate in space. Besides, we also conducted a comparison between the BGK model and MRT model, and found that the
present MRT model could be more accurate and more stable than BGK model through tuning the relaxation parameters properly. And finally, we would also like to point out that, based on the superiority of the MRT model and the role of the
NACDEs in describing the physical phenomena caused by the convection and diffusion processes, the present
work may promote the MRT model in the study of heat and mass transfer \cite{Succi, Guo2013}, multiphase flows and crystal growth based on phase-field models \cite{Zheng, Liang, Boettinger}.

\section*{Acknowledgments}

The authors acknowledge support from the National Natural
Science Foundation of China (Grants No. 11272132 and No. 51125024).

\end{document}